\documentclass[final,1p,times,number]{elsarticle}

\usepackage{amsmath}
\usepackage{amssymb}
\usepackage{amsfonts}
\usepackage{graphicx}
\usepackage{epstopdf}
\usepackage{color}
\usepackage{bm}
\usepackage{multirow}
\usepackage{natbib}
\usepackage{hyperref}
\usepackage{xcolor}
\usepackage{comment}

\usepackage{ifpdf}
\ifpdf%
\usepackage{pdflscape}
\else
\usepackage{lscape}
\fi

\usepackage{ifxetex}
\ifxetex
\usepackage{fontspec}
\else
\fi

\DeclareMathOperator{\Tr}{Tr}

\DeclareMathOperator{\spp}{sp}

\DeclareMathOperator{\sgn}{sgn}

\DeclareMathOperator{\im}{Im}

\DeclareMathOperator{\re}{Re}
\DeclareMathOperator{\arcsinh}{arcsinh}

\journal{Annals of Physics}
\biboptions{sort&compress}

\begin{document}

\begin{frontmatter}
\title{Multifractally-enhanced superconductivity in thin films}

\author[LITP,HSE]{I. S. Burmistrov\corref{cor}}
\author[KIT,TKM,PTI]{I. V. Gornyi}
\author[KIT,TKM,PNPI,LITP]{A. D. Mirlin}

\address[LITP]{L. D. Landau Institute for Theoretical Physics, Semenova 1a, 142432, Chernogolovka, Russia}
\address[HSE]{\mbox{Laboratory for Condensed Matter Physics, National Research University Higher School of Economics, 101000 Moscow, Russia}}

\address[KIT]{Institute for Quantum Materials and Technologies,  Karlsruhe Institute of Technology, 76021 Karlsruhe, Germany}
\address[TKM]{\mbox{Institut f\"ur Theorie der Kondensierten Materie, 76128 Karlsruhe, Germany}}
\address[PTI]{Ioffe Institute, 194021 St.~Petersburg, Russia}
\address[PNPI]{Petersburg Nuclear Physics Institute, 188300 St. Petersburg, Russia}
\cortext[cor]{Corresponding author. Fax: \texttt{+7-495-702-9317} e-mail: burmi@itp.ac.ru}

\begin{abstract}
The multifractal superconducting state originates from the interplay of Anderson localization and interaction effects. In this article we overview the recent theory of the superconductivity enhancement by multifractality and extend it to describe the spectral properties of superconductors on the scales of the order of the superconducting gap. Specifically, using the approach based on renormalization group within the nonlinear sigma model, we develop the theory of a multifractal superconducting state in thin films. We derive a modified Usadel equation that incorporates the interplay of disorder and interactions at energy scales larger than the spectral gap and study the effect of such an interplay on the low-energy physics. We determine the spectral gap at zero temperature which occurs to be proportional to the multifractally enhanced superconducting transition temperature. 
The modified Usadel equation results in the disorder-averaged density of states that, near the spectral gap, resembles the one obtained in the model of a spatially random superconducting order parameter. 
We reveal strong mesoscopic fluctuations of the local density of states in the superconducting state. 
Such strong mesoscopic fluctuations imply that the 
interval
of energies 
in which
the superconducting gap establishes is parametrically large in systems with multifractally-enhanced superconductivity.
\end{abstract}


\end{frontmatter}

\section{Introduction}

Anderson localization \cite{Anderson1958} and superconductivity are two fundamental quantum phenomena. Semiclassicaly, i.e., without taking into account the quantum interference effects ({\color{orange}} and also discarding the influence of disorder on interaction\color{black}), superconductivity is not affected by the electron scattering on non-magnetic disorder that is compatible with the symmetry of the superconducting order parameter (so-called ``Anderson theorem'' \cite{Gor'kovAbrikosov1959a,Gor'kovAbrikosov1959b,Anderson1959}). However, at the next level of sophistication -- with quantum effects included, the Anderson theorem meets Anderson localization. 
At this level, superconductivity and Anderson localization were considered as antagonists: strong localization \cite{Sadovskii1984,Ma1985,Kapitulnik1985,Kapitulnik1986} or interplay of weak disorder and Coulomb interaction \cite{Maekawa1981,Takagi1982,Maekawa1984,Anderson1983,Castellani1984,Bulaevskii1985,Finkelstein1987,KB1993,KB1994,Finkelstein1994} were predicted to destroy superconductivity \footnote{{\color{orange}} It is worth mentioning that in Refs. \cite{Sadovskii1984,Ma1985,Kapitulnik1985,Kapitulnik1986} there was an interval of disorder in which superconductivity and  localization coexist. However, superconductivity was monotonically suppressed with increasing disorder.}. 
This view was supported by a discovery \cite{Goldman1989} and the further study  of the superconductor-insulator transition (SIT) in thin films (see Refs. \cite{Goldman1998,Gantmakher2010,Sacepe2020} for a review). 

The paradigm of suppression of superconductivity by Anderson localization has been challenged in Refs. \cite{Feigel'man2007,Feigel'man2010}, where the enhancement of the superconducting transition temperature, $T_c$, due to the multifractal behavior of wave functions near the Anderson transition (e.g., in three-dimensional disordered systems) 
has been demonstrated for the situation when the long-ranged Coulomb repulsion between electrons is not effective. Later, 
the multifractal enhancement of $T_c$ in thin superconducting films has been predicted by the present authors \cite{BGM2012,BGM2015}. Recently, these theoretical predictions have been supported by numerical solution of the disordered attractive Hubbard model on a two-dimensional \cite{Andersen2018} {\color{orange}} and three-dimensional \cite{Garcia2020} \color{black} lattices. Recently, an enhancement of the superconducting transition temperature by disorder observed in monolayer niobium dichalcogenides has been explained by the multifractality \cite{MultifractalExp1,MultifractalExp2}.

The hallmark of the multifractally-enhanced superconductivity in disordered systems is strong mesoscopic fluctuations of the local order parameter and the local density of states \cite{Feigel'man2010}. The point-to-point fluctuations  of the local density of states have been observed in many experiments on tunneling spectroscopy on thin superconducting films \cite{Sacepe2008,Sacepe2010,Sacepe2011,Sherman2014,Mondal2011,Noat2013}.
Recently, these fluctuations have been retrieved from numerical solution of disordered attractive two-dimensional Hubbard model \cite{Fan2020,Stosiek2020}.

For temperatures above the critical temperature, $T>T_c$, the point-to-point fluctuations of the local density of states observed in the tunneling experiments are in qualitative agreement with the theory \cite{BGM2016} developed by the present authors for the mesoscopic fluctuations of the local density of states in the normal phase of disordered superconducting films. 
Our theory employs the renormalization-group (RG) framework within the non-linear sigma-model (NLSM) formalism. The advantage of such approach is the possibility to take into account mutual influence of disorder and interactions in all channels (particle-hole as well as particle-particle) and to describe systems with short-range and Coulomb interaction on equal footing. 

The experimental data on the fluctuations of the local density of states below $T_c$, as well as recent numerical data on the disordered Hubbard model represent a new challenge for an analytical theory. In the present paper, we develop a theory of the mesoscopic fluctuations of the local density of states in the multifractal superconducting state. For the sake of concreteness, we shall focus on the case of thin superconducting films. We assume the absence of long-range (Coulomb) interaction, which can be achieved in structures put on a substrate with a high dielectric constant, and consider the regime of an intermediate disorder strength \cite{BGM2012,BGM2015}, in which the superconducting transition temperature is parametrically enhanced by multifractality compared to the conventional Bardeen-Cooper-Schrieffer (BCS) result. We calculate the average density of states and its mesoscopic fluctuations in the low-temperature limit in the presence of the interplay between disorder and interactions. 

We demonstrate that the gap in the disorder-averaged density of states at zero temperature is proportional to the superconducting transition temperature, i.e., it is also enhanced by multifractality. In spite of the enhancement of the spectral gap, we find that the combined effect of disorder and interaction results in the suppression of the coherence peaks. Their height is finite and controlled by the ratio of a bare disorder and interaction. We also show that the mesoscopic fluctuations of the local density of states in the superconducting state are enhanced in the same way as in the normal phase for temperatures above $T_c$. 

On the technical side, in order to describe the superconducting state, we derive the Usadel equation modified by renormalization effects due to interplay of disorder and interactions (in both particle-hole and particle-particle channels) at short scales. This modified Usadel equation can be reformulated as a standard Usadel equation, but with an energy dependent gap function, and yields the same estimate for the superconducting transition temperature as the one derived by means of the renormalization group in the normal phase. We discuss the relation between the approaches based on the renormalization of the sigma model and the self-consistent solution of the gap equation in the presence of multifractal correlations of matrix elements.

The structure of the paper is as follows. In Sec. \ref{s2} we overview the formalism of the Finkel'stein NLSM as applied to superconducting systems \cite{Finkelstein1987,Finkelstein1994,BGM2012,BGM2015}. The mean-field description of the superconducting state is briefly outlined in Sec. \ref{s3}. The description of the superconducting state beyond the mean-field approximation is developed in Sec. \ref{s4}. The developed theory is used to estimate the superconducting transition temperature (Sec. \ref{sec:Tc}) and the energy dependence of the gap function (Sec. \ref{sec:GF}). The results for the local density of states and its mesoscopic fluctuations in the superconducting state are presented in Sec. \ref{Sec:AvDOS} and Sec. \ref{sec:MesoLDOS}, respectively. Finally, our results and conclusions are summarized in Sec. \ref{s5}.

\section{Formalism of Finkel'stein NLSM\label{s2}}

\subsection{Finkel'stein NLSM action in the normal state}

The action of the Finkel'stein NLSM is given as the sum of the non-interacting NLSM, $S_\sigma$, 
and contributions resulting from electron-electron interactions, $S_{\rm int}^{(\rho)}$ (the particle-hole singlet channel), 
$S_{\rm int}^{(\sigma)}$ (the particle-hole triplet channel), and $S_{\rm int}^{(c)}$ (the particle-particle channel) (see Refs. \cite{Fin,KB,Burmistrov2019} for a review):
\begin{gather}
S=S_\sigma + S_{\rm int}^{(\rho)}+S_{\rm int}^{(\sigma)}+S_{\rm int}^{(c)} ,
\label{eq:NLSM}
\end{gather}
where
\begin{subequations}
\begin{align}
S_\sigma & = -\frac{g}{32} \int d\bm{r} \Tr (\nabla Q)^2 + 2 Z_\omega \int d\bm{r} \Tr \hat \varepsilon Q ,
\label{Ss} \\
S_{\rm int}^{(\rho)} & =-\frac{\pi T}{4} \Gamma_s \sum_{\alpha,n} \sum_{r=0,3}
\int d\bm{r} \Tr I_n^\alpha t_{r0} Q \Tr I_{-n}^\alpha t_{r0} Q,
\label{Srho}\\
S_{\rm int}^{(\sigma)}& =-\frac{\pi T}{4} \Gamma_t \sum_{\alpha,n} \sum_{r=0,3}\sum_{j=1,2,3}
\int d\bm{r} \Tr I_n^\alpha t_{rj} Q \Tr I_{-n}^\alpha t_{rj} Q ,
\label{Ssigma}\\
S_{\rm int}^{(c)}& = -\frac{\pi T}{4}  \Gamma_c \sum_{\alpha,n} \sum_{r=1,2}  \int d\bm{r} \Tr  t_{r0} L_n^\alpha Q \Tr t_{r0} L_n^\alpha Q  . 
\label{Sc}
\end{align}
\end{subequations}
Here, $Q(\bm{r})$ is the matrix field in the replica, Matsubara, spin, and particle-hole spaces. The trace $\Tr$ acts in the same spaces. The matrix field obeys the nonlinear constraint, as well as charge-conjugation symmetry relation:
\begin{gather}
Q^2(\bm{r})=1, \quad \Tr Q = 0, \qquad Q=Q^\dag = -C Q^T C ,\qquad C=i t_{12} \ .
\label{eq:constraints}
\end{gather}
The nonlinear constraint on the matrix field $Q$ suggests the following parametrization:
\begin{equation}
Q = T^{-1}\Lambda T, \quad T^\dag=T^{-1}, \quad C T^T  =  T^{-1} C , \qquad\Lambda_{nm}^{\alpha\beta}  = \sgn \varepsilon_n \, \delta_{\varepsilon_n,\varepsilon_m} \delta^{\alpha\beta}t_{00} ,
\label{eq:rep:T}
\end{equation}
where $\alpha,\beta = 1,\dots, N_r$ stand for replica indices and integers $n,m$ correspond to the Matsubara fermionic frequencies $\varepsilon_n = \pi T (2n+1)$. 
The representation \eqref{eq:rep:T} implies that the diffusive fluctuations in the normal state are associated with smooth variations around the metallic saddle point $Q=\Lambda$. 

The action \eqref{eq:NLSM} contains three constant matrices:
\begin{equation}
\hat \varepsilon_{nm}^{\alpha\beta} =\varepsilon_n \, \delta_{\varepsilon_n,\varepsilon_m}\delta^{\alpha\beta} t_{00}, \quad 
(I_k^\gamma)_{nm}^{\alpha\beta} =\delta_{\varepsilon_n-\varepsilon_m,\omega_k}\delta^{\alpha\beta}\delta^{\alpha\gamma} t_{00}, \quad
(L_k^\gamma)_{nm}^{\alpha\beta} =\delta_{\varepsilon_n+\varepsilon_m,\omega_k}\delta^{\alpha\beta}\delta^{\alpha\gamma} t_{00} \ .
\end{equation}
The sixteen matrices,
\begin{equation}
\label{trj}
t_{rj} = \tau_r\otimes s_j, \qquad r,j = 0,1,2,3  ,
\end{equation}
operate in the particle-hole (subscript $r$) and spin (subscript $j$) spaces. The matrices 
$\tau_0, \tau_1, \tau_2, \tau_3$ and $s_0, s_1, s_2, s_3$ are the standard sets of the Pauli matrices.

The bare value of the coupling constant $g$ coincides with the Drude value of the conductance (in units $e^2/h$ and including spin). The parameter $Z_\omega$ describes the frequency renormalization upon the renormalization group flow \cite{Fin}. The bare value of $Z_\omega$ is equal to $\pi \nu/4$, where $\nu$ denotes the bare density of states at the Fermi level including a spin-degeneracy factor. The interaction amplitude $\Gamma_s$ ($\Gamma_t$) encodes interaction in the singlet (triplet) particle-hole channel. The interaction in the Cooper channel is denoted as $\Gamma_c$. Its negative magnitude, $\Gamma_c<0$, corresponds to an attraction in the particle-particle channel. In what follows, it is convenient to introduce the dimensionless interaction parameters, $\gamma_{s,t,c} = \Gamma_{s,t,c}/Z_\omega$. 
If Coulomb interaction is present, the following relation holds: $\gamma_s=-1$. This condition remains intact under the action of the renormalization group flow \cite{Fin,BGM2015}. In the case of the BCS model and, under an assumption of strong disorder, 
$\omega_D\tau\ll 1$ ($\omega_D$ is the Debye frequency),
 the bare values of interaction amplitudes are related as $-\gamma_{s0}=\gamma_{t0}=\gamma_{c0}$. 
In what follows, we shall refer to such a relation between interaction amplitudes as the \textit{BCS line}.

\subsection{Finkel'stein NLSM action in the superconducting state}

The symmetry breaking that leads to the appearance of the superconducting state 
changes the saddle point of the NLSM \cite{FLS2000,YL2001,LK2007,Konig2015}. We shall follow the approach of Ref. \cite{Konig2015} which allows us to take  into account the effects of disorder and residual quasiparticle interactions beyond the mean-field level. \color{black}

Let us single out the term with $n=0$ from the expression for $S_{\rm int}^{(c)}$, Eq. \eqref{Sc}. Then, upon the Hubbard--Stratonovich decoupling by spatially dependent fields $\Delta_r^\alpha(\bm{r})$, $r=1,2$, we find 
\begin{equation}
S_{\rm int}^{(c)} =
\frac{4 Z_\omega}{\pi T \gamma_c} \int d\bm{r} \sum_{\alpha}\sum_{r=1,2} 
\bigl [ \Delta_r^\alpha(\bm{r}) \bigr ]^2
+ 2 Z_\omega \int d\bm{r} \sum_{\alpha}\sum_{r=1,2}  
 \Delta_r^\alpha(\bm{r}) \Tr t_{r0} L_0^\alpha Q
  + \tilde S_{\rm int}^{(c)} \ ,
  \label{eq:Sc:HS}
 \end{equation} 
 where
\begin{equation}
 \tilde S_{\rm int}^{(c)} 
=-\frac{\pi T}{4}  \Gamma_c \sum_{\alpha,n\neq 0} \sum_{r=1,2}  \int d\bm{r}\ \Bigl (\Tr   t_{r0} L_n^\alpha Q \Bigr )^2
 \ .
\end{equation}
Since, by construction, the action depends quadratically on $\Delta_r^\alpha(\bm{r})$, we can solve the equation of motion instead of performing the functional integral. The equation of motion reads
\begin{equation}
\Delta_r^\alpha(\bm{r}) = \frac{\pi T}{4} |\gamma_c| \Tr t_{r0} L_0^\alpha Q(\bm{r}) , \qquad r=1,2 .
\label{eq:SCE:full}
\end{equation}
We note that the charge-conjugation symmetry, Eq. \eqref{eq:constraints}, guarantees that $\Delta_{1,2}^\alpha(\bm{r})$ are real functions. 

The presence of nonzero $\Delta_r^\alpha(\bm{r})$ changes the saddle-point equation for $Q$. Performing variation of the NLSM action \eqref{eq:NLSM} with $S_{\rm int}^{(c)}$ replaced by $\tilde S_{\rm int}^{(c)}$, we derive the following saddle-point equation:
\begin{gather}
- D\nabla (Q\nabla Q) +[\hat\varepsilon,  Q] 
+\sum_{\alpha} \sum_{r=1,2}  
 \Delta_r^\alpha(\bm{r}) [t_{r0} L_0^\alpha, Q] 
 -\frac{\pi T}{4} \sum_{\alpha,n} \sum_{r=0,3}\sum_{j=0}^3 \gamma_j 
[I_{-n}^\alpha t_{rj}, Q] \Tr I_n^\alpha t_{rj} Q 
\notag \\
- \frac{\pi T\gamma_c}{4}  \sum_{\alpha,n\neq 0} \sum_{r=1,2}  [t_{r0} L_n^\alpha, Q] \Tr  t_{r0} L_n^\alpha Q 
 =0 ,
\label{eq:Usadel:full:0} 
\end{gather}
Here, $\gamma_0\equiv \gamma_s$ and $\gamma_1=\gamma_2=\gamma_3\equiv \gamma_t$,  and we have introduced the diffusion coefficient $D=g/(16 Z_\omega)$. We mention that in the absence of interaction Eq. \eqref{eq:Usadel:full:0} is nothing but the Usadel equation, well known in the theory of dirty superconductors.

\color{black}
Substitution of $\Delta_r^\alpha(\bm{r})$ from the self-consistency equation \eqref{eq:SCE:full} into the Usadel equation \eqref{eq:Usadel:full:0} results in the nonlinear matrix equation for $Q(\bm{r})$. In general, such equation 
has many spatially dependent solutions for $Q(\bm{r})$ which mimic fluctuations of a disorder potential in the original microscopic formulation of the problem. To enumerate all these solutions and to perform summation over them seem like a daunting task. In order to circumvent this difficulty we shall use the smallness of the bare resistance, $1/g\ll 1$. Then, we can sum over spatially dependent solutions for $Q(\bm{r})$ by treating the fluctuations around the spatially independent solution of the Usadel equation by means of the renormalization group. 
 We start from splitting 
 \color{black}
 $\Delta_r^\alpha$ into constant and spatially dependent part,
\begin{equation}
\Delta_r^\alpha(\bm{r})  = \underline{\Delta}_r^\alpha+\delta \Delta_r^\alpha(\bm{r}),
\qquad  \int d\bm{r} \ \delta \Delta_r^\alpha(\bm{r}) = 0 .
\end{equation}
The spatially varying part, $\delta \Delta_r^\alpha(\bm{r})$, contains information about the mesoscopic fluctuations of the superconducting order parameter. These mesoscopic fluctuations turn out to be large (see \ref{App:MesoDelta}) \cite{Skvortsov2005,Feigel'man2010,Fan2020,Stosiek2020}. 
In order to take them into account, we perform a formally exact integration over $\delta \Delta_r^\alpha(\bm{r})$. As a result, the spatial fluctuations
of the Hubbard-Stratonovich field corresponding to the order parameter get fully encoded in additional correlations of the $Q$ field.  
This procedure results in a modification of  $S_{\rm int}^{(c)}$,
\begin{equation}
S_{\rm int}^{(c)} =
\frac{4 Z_\omega V}{\pi T \gamma_c}  \sum_{\alpha}\sum_{r=1,2} 
\bigl [ \underline{\Delta}_r^\alpha \bigr ]^2
+ 2 Z_\omega  V \sum_{\alpha}\sum_{r=1,2}  
\underline{\Delta}_r^\alpha \Tr t_{r0} L_0^\alpha \overline{Q}
  + \hat S_{\rm int}^{(c)} ,
  \end{equation}
 where $V$ denotes the volume of a superconductor,
 \begin{equation}
 \overline{Q} = \frac{1}{V} \int d\bm{r} Q(\bm{r}) ,
 \end{equation}
   and 
 \begin{equation}
 \hat S_{\rm int}^{(c)} 
=  -\frac{\pi T}{4}  \Gamma_c \sum_{\alpha,n} \sum_{r=1,2}  \int d\bm{r}\left [ \Tr  t_{r0} L_n^\alpha \Bigr (Q-\overline{Q} \delta_{n,0} \Bigr ) \right ]^2\ .
\label{eq:hatSc}
\end{equation}
After integrating over $\delta \Delta_r^\alpha(\bm{r})$, the saddle-point equation \eqref{eq:Usadel:full:0} gets modified, 
\begin{gather}
- D\nabla (Q\nabla Q) +[\hat\varepsilon,  Q] 
+[\hat{\underline{\Delta}}, Q] 
 -\frac{\pi T}{4} \sum_{\alpha,n} \sum_{r=0,3} \sum_{j=0}^3 \gamma_j 
[I_{-n}^\alpha t_{rj}, Q] \Tr I_n^\alpha t_{rj} Q 
\notag \\
- \frac{\pi T\gamma_c}{4}  \sum_{\alpha,n} \sum_{r=1,2}  [t_{r0} L_n^\alpha, Q] \Tr  t_{r0} L_n^\alpha (Q -\overline{Q}\delta_{n,0})
 =0 ,
\label{eq:Usadel:full} 
\end{gather}
where 
\begin{equation}
\hat{\underline{\Delta}}=\sum_{\alpha} \sum_{r=1,2}  
 \underline{\Delta}_r^\alpha t_{r0} L_0^\alpha , \qquad 
 \underline{\Delta}_r^\alpha = \frac{\pi T}{4} |\gamma_c| \Tr t_{r0} L_0^\alpha \overline{Q} \ .
 \label{eq:SCE:full:1} 
\end{equation}
It is worth noting that the information about the disorder-induced spatial fluctuations of the Hubbard-Stratonovich field 
$\Delta_r^\alpha$, and, hence, about the spatial fluctuations of the order parameter, 
is not lost after the integration over $\delta \Delta_r^\alpha(\bm{r})$ and remains encoded in Eq.~\eqref{eq:Usadel:full}
through the coupling of $\hat{\underline{\Delta}}$ and $Q$ and the fluctuations of field $Q$.
Before considering the effect of such fluctuations, in the following section we analyse the mean-field solution
of the modified Usadel equation.

\section{Mean-field description of the superconducting state\label{s3}}

Let us seek the solution of the Usadel equation \eqref{eq:Usadel:full} and the self-consistency equation \eqref{eq:SCE:full:1}  in the following form
\begin{gather}
\underline{Q}_{nm}^{\alpha\beta} = \Bigl (\cos\theta_{\varepsilon_n} 
\sgn \varepsilon_n \delta_{\varepsilon_n,\varepsilon_m}
 +  t_\phi \sin \theta_{\varepsilon_n} \delta_{\varepsilon_n,-\varepsilon_m} \Bigr )\  \delta^{\alpha\beta}, \qquad t_\phi =\cos\phi \ t_{10}+\sin\phi\ t_{20} , \notag \\
 \underline{\Delta}_1^\alpha=\Delta \cos\phi, \qquad \underline{\Delta}_2^\alpha =\Delta \sin \phi .
 \label{eq:sol:sp}
\end{gather}
Here, we assume that the spectral angle $\theta_{\varepsilon_n}$ is an even function of $\varepsilon_n$.
We note that the above ansatz, $\underline{Q}$, satisfies the charge-conjugation condition. 
Substituting Eq. \eqref{eq:sol:sp} into the Usadel equation  \eqref{eq:Usadel:full}, we obtain 
\begin{equation}
\frac{D}{2} \nabla^2 \theta_{\varepsilon_n} - |\varepsilon_n|\sin \theta_{\varepsilon_n}+\Delta \cos\theta_{\varepsilon_n}
=0 .
\label{eq:Usadel:theta}
\end{equation}
The self-consistency equation \eqref{eq:SCE:full:1} transforms into the following relation:
\begin{equation}
\Delta = \pi  T |\gamma_c|\sum_{\varepsilon_n} \sin \theta_{\varepsilon_n} .
\label{eq:SCE:theta}
\end{equation}
{\color{orange}} Since the superconductor order parameter $\Delta$ is spatially independent by construction, \color{black}
we consider a spatially independent solution for $\theta_{\varepsilon_n}$. 
Then, we find:
\begin{equation}
\cos\theta_{\varepsilon_n} = \frac{|\varepsilon_n|}{\sqrt{\varepsilon_n^2+\Delta^2}},\qquad
\sin \theta_{\varepsilon_n} = \frac{\Delta}{\sqrt{\varepsilon_n^2+\Delta^2}}\ ,
\end{equation}
where $\Delta$ satisfies the self-consistency relation
\begin{equation}
\Delta = 
\pi  T |\gamma_c|\sum_{\varepsilon_n} \frac{\Delta}{\sqrt{\varepsilon_n^2+\Delta^2}} \ .
\label{eq:sce:bcs}
\end{equation}
This is nothing but the standard self-consistency condition for the gap $\Delta$ in the BCS theory \cite{BCS1957}.

The saddle-point solution \eqref{eq:sol:sp} can be conveniently written as rotation around the matrix $\Lambda$,
\begin{gather}
\underline{Q} = R^{-1}\Lambda R, \qquad R_{nm}^{\alpha\beta} =
\Bigl [ \delta_{\varepsilon_n,\varepsilon_m}
\cos (\theta_{\varepsilon_n}/2) -  t_{\phi} \delta_{\varepsilon_n,-\varepsilon_m} \sgn \varepsilon_m \sin (\theta_{\varepsilon_n}/2) \Bigr] \delta^{\alpha\beta} \ . \\
(R^{-1})_{nm}^{\alpha\beta} =\Bigl [ \delta_{\varepsilon_n,\varepsilon_m}
\cos (\theta_{\varepsilon_n}/2) - t_{\phi} \delta_{\varepsilon_n,-\varepsilon_m} \sgn \varepsilon_n \sin (\theta_{\varepsilon_n}/2) \Bigr ] \delta^{\alpha\beta}\ . \notag
\end{gather}
We note that the matrix $R$ satisfies the relation, $C R^T=R^{-1} C$.
We emphasize that although the spatially independent saddle point ansatz \eqref{eq:sol:sp} solves Usadel equation \eqref{eq:Usadel:full}, the solution \eqref{eq:sol:sp} is completely insensitive to disorder and residual interaction between quasiparticles in accordance with the ``Anderson theorem''  \cite{Gor'kovAbrikosov1959a,Gor'kovAbrikosov1959b,Anderson1959}.

\section{Beyond the mean-field approximation \label{s4}}

In order to see the effect of disorder and residual interactions between quasiparticles, one needs to go beyond the mean-field approximation of the previous section. The fluctuations of the matrix $Q$ around the saddle-point ansatz \eqref{eq:sol:sp} modify the effective potential for the spectral angle $\theta_\varepsilon$. In what follows, we establish a perturbative renormalization group approach for accounting these fluctuations.

\subsection{Perturbative expansion}

Taking into account the fluctuations of $Q$, we renormalize the NLSM action. For this purpose, we shall develop a peturbation expansion around the saddle point $\underline{Q}$,  using the square-root parametrization of the matrix field $Q$:
\begin{gather}
Q = R^{-1} \Bigl( W +\Lambda \sqrt{1-W^2}\Bigr ) R, \qquad W= \begin{pmatrix}
0 & w\\
\overline{w} & 0
\end{pmatrix} .
\label{eq:Q-W}
\end{gather}
We adopt the following notations: $W_{n_1n_2} = w_{n_1n_2}$ and $W_{n_2n_1} = \overline{w}_{n_2n_1}$, where $n_1\geqslant 0$ and $n_2< 0$.
The blocks $w$ and $\overline{w}$ satisfy the charge-conjugation constraints:
\begin{gather}
\overline{w} = -C w^T C,\qquad w = - C w^* C .
\end{gather}
These constraints imply that some elements $(w^{\alpha\beta}_{n_1n_2})_{rj}$ in the expansion
$w^{\alpha\beta}_{n_1n_2}= \sum_{rj} (w^{\alpha\beta}_{n_1n_2})_{rj} t_{rj}$ are purely real and the others are purely imaginary.

In this paper, we restrict our considerations to the expansion of the renormalization group equations to the lowest order in residual electron-electron interactions. This is justified in the case of weak short-ranged interaction, which corresponds to small magnitudes of the bare interaction parameters, $|\gamma_{s0}|, |\gamma_{t0}|, |\gamma_{c0}| \ll  1$. Then, in order to study the effect of fluctuations, we shall need the propagators for diffusive modes in the noninteracting theory. In other words, the propagators are determined by the NLSM action in which terms $S_{\rm int}^{(\rho)}$, $S_{\rm int}^{(\sigma)}$, and $\hat S_{\rm int}^{(c)}$ are omitted. Within this perturbative-in-interaction scheme, one finds 
\begin{gather}
\Bigl \langle [w_{rj}(\bm{p})]^{\alpha_1\beta_1}_{n_1n_2} [\overline{w}_{rj}(-\bm{p})]^{\beta_2\alpha_2}_{n_4n_3} 
\Bigr \rangle =  \frac{2}{g} \delta^{\alpha_1\alpha_2} \delta^{\beta_1\beta_2}
\delta_{\varepsilon_{n_1},\varepsilon_{n_3}} 
\delta_{\varepsilon_{n_2},\varepsilon_{n_4}}
\mathcal{D}_p(i\varepsilon_{n_1},i\varepsilon_{n_2}) ,\quad  
r,j=0,\dots, 3, \notag \\
\mathcal{D}_p(i\varepsilon_{n_1},i\varepsilon_{n_2}) 
=\frac{1}{D p^2+ E_{\varepsilon_{n_1}}+E_{\varepsilon_{n_2}}}  , 
\label{eq:prop:diff}
\\ 
E_{\varepsilon_{n}} = |\varepsilon_{n}| \cos \theta_{\varepsilon_{n}} +
\Delta\sin \theta_{\varepsilon_{n}} \ .
\label{eq:E-eps}
\end{gather}
We note that we have not yet fixed $\theta_{\varepsilon_{n}}$ at this stage.

\subsection{Renormalization of the NLSM action}

The interaction of the diffusive modes encoded in  $W$ renormalizes the NLSM action. 
Since we consider the spatially independent saddle point, $\underline{Q}$, and $S_{\rm int}^{(\rho)}$, $S_{\rm int}^{(\sigma)}$, and $\hat S_{\rm int}^{(c)}$ are zero at this saddle point, we are interested in modifications of these terms in 
the NLSM action by the quantum corrections that can be expressed through the diffusive propagators \eqref{eq:prop:diff}. 
To the lowest order in disorder it is enough to approximate $Q$ as 
$$Q\to \underline{Q}+R^{-1} W R.$$ 
This results in
\begin{align}
S_{\rm int}^{(\rho)}+S_{\rm int}^{(\sigma)} 
\to  -\frac{\pi T}{4}\int d \bm{r} 
 \left \langle  \sum_{\alpha,n}\sum_{r=0,3}\sum_{j=0}^3 \Gamma_j  \Tr \bigl [ R I_n^\alpha t_{rj} R^{-1} W\bigr ]  
\Tr \bigl [ R I_{-n}^\alpha t_{rj} R^{-1} W \bigr ] \right \rangle .
\end{align}
Then, using Eq. \eqref{eq:prop:diff}, we retrieve
\begin{gather}
S_{\rm int}^{(\rho)}+S_{\rm int}^{(\sigma)} 
= -\frac{32 \pi T N_r V}{g} 
\sum_{n_1n_2} 
\int \frac{d^d\bm{q}}{(2\pi)^d}\mathcal{D}_q(i\varepsilon_{n_1},i\varepsilon_{n_2})
\notag \\
\times \sum_n 
\Biggl\{ 
(\Gamma_s+ 3\Gamma_t) \delta_{\varepsilon_{n_1}+\varepsilon_{n_2},\omega_n}
+(\Gamma_s+ 3\Gamma_t)\bigl (\delta_{\varepsilon_{n_1}-\varepsilon_{n_2},\omega_n}-\delta_{\varepsilon_{n_1}+\varepsilon_{n_2},\omega_n}\bigr )
\Biggl [\cos^2 \left (\theta_{\varepsilon_{n_1}}/2\right )
 \notag \\
\times \cos^2 \left (\theta_{\varepsilon_{n_2}}/2\right )
+ \sin^2 \left (\theta_{\varepsilon_{n_1}}/2\right )
\sin^2 \left (\theta_{\varepsilon_{n_2}}/2\right )
\Biggr ]
-(\Gamma_s-3\Gamma_t)  \delta_{\varepsilon_{n_1}+\varepsilon_{n_2},\omega_n} \sin \theta_{\varepsilon_{n_1}}\sin \theta_{\varepsilon_{n_2}}
\Biggr \} 
\notag \\
\to
 \frac{32\pi T N_r V}{g} (\Gamma_s-3\Gamma_t) 
 \sum_{\varepsilon,\varepsilon^\prime>0} \sin \theta_{\varepsilon} \sin \theta_{\varepsilon^\prime} 
 \int \frac{d^d\bm{q}}{(2\pi)^d}\mathcal{D}_q(i\varepsilon,-i\varepsilon^\prime)\ .
\label{eq:ren:Srho+sigma}
\end{gather}
Here, in the last line, we omitted the term proportional to $\Gamma_s+3\Gamma_t$ that involves $\theta_{\varepsilon_n}$ in the diffusion propagator only, as it yields a negligible correction with respect to the term we retain. In a similar way, one can renormalize the interaction in the Cooper channel,
\begin{gather}
\hat{S}_{\rm int}^{(c)} \to  -
  \frac{\pi T \Gamma_c}{4} \sum_{\alpha,n} \!\sum_{r=1,2}  
\int\!\! d\bm{r} \left \langle \left (\Tr \bigl [ R t_{r0} L_n^\alpha R^{-1} W \bigr ] \right )^2 \right \rangle 
+  \frac{\pi T\Gamma_c}{4 V}\!  \sum_{\alpha}\! \sum_{r=1,2} \!\left \langle  
 \left (\int\!\! d\bm{r} \Tr \bigl [ R t_{r0} L_n^\alpha R^{-1} W \bigr ] \right )^2 \right \rangle 
\notag \\
= -\frac{64 \pi T \Gamma_c V N_r}{g} \sum_{n_1,n_2} 
\int \frac{d^d\bm{q}}{(2\pi)^d}\mathcal{D}_q(i\varepsilon_{n_1},i\varepsilon_{n_2}) \sum_{n}
\Biggl \{\delta_{\varepsilon_{n_1}-\varepsilon_{n_2},\omega_{n}}
+\bigl (\delta_{\varepsilon_{n_1}+\varepsilon_{n_2},\omega_{n}}
-\delta_{\varepsilon_{n_1}-\varepsilon_{n_2},\omega_{n}}\bigr )
\notag \\
\times
\Biggl [\cos^2 \left (\theta_{\varepsilon_{n_1}}/2\right )
\cos^2 \left (\theta_{\varepsilon_{n_2}}/2\right )
+ \sin^2 \left (\theta_{\varepsilon_{n_1}}/2\right )
\sin^2 \left (\theta_{\varepsilon_{n_2}}/2\right )
\Biggr ]\Bigl [ 1- (2\pi)^d \delta(\bm{q}) \delta_{n,0}/V\Bigr ]
\Biggr \} 
\notag \\
\to -\frac{32 \pi T \Gamma_c N_r}{g} \sum_{\varepsilon>0}  \mathcal{D}_{q=0}(i\varepsilon,-i\varepsilon) \sin^2 \theta_\varepsilon .
\label{eq:ren:Scooper1}
\end{gather}
As above, we have neglected here the terms that involve $\theta_{\varepsilon_n}$ in the diffusion propagator only. As one can see, the renormalization of $\hat{S}_{\rm int}^{(c)}$ results in the non-extensive term, proportional to $V^0$. In what follows, we shall safely neglect this term in the thermodynamic limit $V\to \infty$.

We mention that the above computation is similar to the background field renormalization in the normal state but with the specific slow field $\underline{Q}$ (see Supplemental Materials  of Ref. \cite{BGM2012} and Appendix  A of Ref. \cite{BGM2015} for details). We also note that to the lowest order in interactions $\gamma_{s,t,c}$ the renormalization of the parameter $Z_\omega$ coincides with the renormalization of the $Q$ matrix, i.e., with the $Z$ factor that determines the renormalization of the average density of states.

Equations \eqref{eq:ren:Srho+sigma} -- \eqref{eq:ren:Scooper1} determine the correction 
$\delta  S_{\rm fl}[\underline{Q}]$ to the classical action  due to quantum fluctuations (treated to the lowest order in $1/g$). 
Including this correction, we obtain the perturbatively modified action in the following form: 
\begin{align}
S[\underline{Q}] + \delta S_{\rm fl}[\underline{Q}]  = & 16\pi T Z_\omega N_r V \Biggl \{
\frac{\Delta^2}{4\pi T \gamma_c}
+
\sum_{\varepsilon>0} \Bigl [ \varepsilon \cos\theta_{\varepsilon} + \Delta \sin \theta_{\varepsilon}\Bigr ] \notag\\
& +\frac{2\pi T(\gamma_s-3\gamma_t)}{g}  \sum_{\varepsilon,\varepsilon^\prime>0} \sin \theta_{\varepsilon} \sin \theta_{\varepsilon^\prime} 
 \int \frac{d^d\bm{q}}{(2\pi)^d}\mathcal{D}_q(i\varepsilon,-i\varepsilon^\prime)
\Biggl \}\ .
\label{eq:eff:pot}
\end{align}
This action will serve as a basis for the modified equations of motion with the quantum interference and interaction effects included.

\subsection{Modified Usadel equation}

Minimizing Eq.~\eqref{eq:eff:pot} with respect to $\theta_\varepsilon$ and $\Delta$, we find ($\varepsilon>0$)
\begin{gather}
-\varepsilon \sin\theta_{\varepsilon} +  \Biggl \{ \Delta
+ 2\pi T \frac{(\gamma_s-3\gamma_t)}{g} \sum_{\varepsilon^\prime>0}  
 \int \frac{d^d\bm{q}}{(2\pi)^d}\Bigl [ \mathcal{D}_q(i\varepsilon,-i\varepsilon^\prime)
+ \mathcal{D}_q(i\varepsilon^\prime,-i\varepsilon) \Bigr ]\sin \theta_{\varepsilon^\prime}\Biggr \}\cos \theta_{\varepsilon}  = 0 \ , \notag \\
\Delta= 2\pi T |\gamma_c| \sum_{\varepsilon^\prime>0}\sin \theta_{\varepsilon^\prime} \ . \label{eq:mod:Usadel:00}
\end{gather}
It is convenient to introduce the energy-dependent quantity 
\begin{equation}
\Delta_{\varepsilon} = -2\pi T  \sum_{\varepsilon^\prime>0}
\Biggl \{\gamma_c - \frac{(\gamma_s-3\gamma_t)}{g} 
 \int \frac{d^d\bm{q}}{(2\pi)^d}\Bigl [ \mathcal{D}_q(i\varepsilon,-i\varepsilon^\prime)
+ \mathcal{D}_q(i\varepsilon^\prime,-i\varepsilon) \Bigr ] \Biggr \}
\sin \theta_{\varepsilon^\prime} \ ,
\label{eq:spectral:gap}
\end{equation}
that we shall term the \textit{gap function}. 
Then, we can write the modified Usadel equation in the following concise form ($\varepsilon>0$),
\begin{equation}
-\varepsilon \sin\theta_{\varepsilon} + \Delta_{\varepsilon} \cos \theta_{\varepsilon}=0 \ .
\label{eq:mod:Usad}
\end{equation}
Solving Eq. \eqref{eq:mod:Usad}, we find the spectral
angle,
\begin{equation}
\sin\theta_{\varepsilon} = \frac{\Delta_{\varepsilon}}{\sqrt{\varepsilon^2+\Delta_{\varepsilon}^2}}, \qquad \cos\theta_{\varepsilon} = \frac{\varepsilon}{\sqrt{\varepsilon^2+\Delta_{\varepsilon}^2}} \ .
\label{eq:sol:theta}
\end{equation}
Supplemented by Eq. \eqref{eq:sol:theta},  Eq. \eqref{eq:spectral:gap} becomes a self-consistent equation for the gap function. We note that, contrary to the gap function, the Hubbard-Stratonovich field (the order parameter) $\Delta$ remains independent of the energy, 
\begin{equation}
\Delta= 2\pi T |\gamma_c| \sum_{\varepsilon>0}  \frac{\Delta_{\varepsilon}}{\sqrt{\varepsilon^2+\Delta_{\varepsilon}^2}} \ .
\label{eq:OP:ME}
\end{equation}

A few remarks are in order here. First, we note that a similar form of the modified Usadel equation has been derived in Refs. \cite{Skvortsov2005,Feigelman2012} in the case of a residual Coulomb interaction and without the exchange interaction, i.e., for $\gamma_s=-1$ and $\gamma_t=0$, by means of the diagrammatic technique \footnote{The Cooperon screening factor of Ref. \cite{Skvortsov2005} is given as $w(\varepsilon)=\Delta_\varepsilon/\Delta$.}. Second, the accuracy of our approximation 
does not allow us to see that $\Delta$ in the definition of $E_{\varepsilon}$, Eq. \eqref{eq:E-eps}, is changed to $\Delta_\varepsilon$.  
However, there is an argument in favor of such transformation from $\Delta$ to $\Delta_\varepsilon$. Let us make a shift  in the spectral angle $\theta_\varepsilon \to \theta_\varepsilon + \delta \theta_\varepsilon$. Then, on one hand, $\delta \theta_\varepsilon$ can be reabsorbed into some $W$ matrix. Therefore, the propagator for  $\delta \theta_\varepsilon$ is related with the propagator for diffusive modes, Eq. \eqref{eq:prop:diff}. On the other hand, the very same propagator can be obtained from the linear variation of the modified Usadel equation. The latter fact implies the presence of $\Delta_\varepsilon$ in the expression for  the propagator of diffusive modes. Therefore, we make the following substitution in the diffusive propagator, 
\begin{equation}
\mathcal{D}^{-1}_q(i\varepsilon,-i\varepsilon^\prime) \to q^2+ L^{-2}_{
\sqrt{\varepsilon^{2}+\Delta_{\varepsilon}^2}+\sqrt{\varepsilon^{\prime 2}+\Delta_{\varepsilon^\prime}^2}
} \ , 
\label{eq:modif:prop:123}
\end{equation}
where $L_{\varepsilon} = \sqrt{D/\varepsilon}$. We note that for energies larger than the spectral gap edge the transformation from $\Delta$ to $\Delta_\varepsilon$
is not essential. 

Finally, the form of the second term in the curly brackets in the right hand side of Eq. \eqref{eq:spectral:gap} coincides exactly with the perturbative correction to the interaction in the Cooper channel originating from the interplay of disorder  and interactions in the particle-hole channel. Since this correction appeared in a procedure similar to the background field renormalization, we are allowed to rewrite Eq. \eqref{eq:spectral:gap}  as the following self-consistency equation for $\Delta_\varepsilon$,
\begin{equation}
\Delta_\varepsilon= - 2\pi T \sum_{\varepsilon^\prime>0} \gamma_c\left (
L_{\sqrt{\varepsilon^{2}+\Delta_{\varepsilon}^2}+\sqrt{\varepsilon^{\prime 2}+\Delta_{\varepsilon^\prime}^2}}
\right )  \frac{\Delta_{\varepsilon^\prime}}{\sqrt{\varepsilon^{\prime 2}+\Delta_{\varepsilon^\prime}^2}}  \ ,
\label{eq:spectral:gap:2}
\end{equation}
where the dependence of $\gamma_c$ on $L$ is governed by the following equation
\begin{equation}
\frac{d \gamma_c}{dy}= -\frac{t}{2} (\gamma_s-3\gamma_t) \ .
\label{eq:gamma:sup}
\end{equation}
Here $y=\ln L/\ell$ and $t=2/(\pi g)$ denotes the dimensionless resistance. 
Comparison of Eq. \eqref{eq:spectral:gap:2} and Eq. \eqref{eq:OP:ME} suggests that the order parameter $\Delta$ coincides with the value of the gap function at energy of the order of $1/\tau$, i.e. $\Delta = \Delta_{\varepsilon\sim 1/\tau}$.

\section{Critical temperature for the transition to the superconducting state \label{sec:Tc}}

The modified self-consistency equation \eqref{eq:spectral:gap:2} allows us to determine the superconducting transition beyond the mean-field approximation. The latter yields the BCS result, 
\begin{equation}
T_c^{\rm BCS} \sim \tau^{-1} \exp(-1/|\gamma_{c0}|),
\label{Tc-BCS}
\end{equation}
see 
Eq. \eqref{eq:sce:bcs}.

\subsection{Estimate for $T_c$ from the renormalization group equations in the normal phase\label{Sec:Tc:RG}}

We start by reminding the reader on how the superconducting transition temperature can be estimated from the renormalization group equations in the normal phase. In the case of weak short-range interactions, these renormalization group equations, to the lowest order in $t$, read
\cite{BGM2012,BGM2015,DellAnna2013}
\begin{subequations}
\begin{align}
\frac{d t}{dy} & = t^2 \bigl ( 1 - \gamma_s/2-3 \gamma_t/2- \gamma_c \bigr ) , \label{eq:rg:final:t}\\
\frac{d\gamma_s}{dy}  & = - \frac{t}{2} \bigl ( \gamma_s+3\gamma_t+2\gamma_c\bigr ), \label{eq:rg:final:gs} \\
\frac{d\gamma_t}{dy}  & = - \frac{t}{2} \bigl (\gamma_s-\gamma_t-2\gamma_c 
 \bigr ), \label{eq:rg:final:gt} \\
\frac{d\gamma_c}{dy} & =  - \frac{t}{2} \bigl (\gamma_s-3\gamma_t\bigr )  - 2\gamma_c^2,
\label{eq:rg:final:gc}
\\
\frac{d\ln Z_\omega}{dy} & = \frac{t}{2} \bigl (\gamma_s+3\gamma_t+2\gamma_c \bigr )  .\label{eq:rg:final:Z}
\end{align}
\end{subequations}

{\color{orange}}
It is worthwhile to mention that, within the lowest order in $t$, the dependence of the renormalization-group functions on the interaction parameters can be computed exactly 
\cite{Fin,KB}. There are subtleties (extensively discussed in the past, see e.g. Ref. \cite{KB1993,KB1994,Finkelstein1994}) related with the renormalization of the Cooper-channel interaction. Recently, the difficulties with the Cooper-channel renormalization have been resolved within the background-field renormalization scheme, and the renormalization-group functions for interaction parameters $\gamma_{s,t,c}$ have  been computed exactly (including dependence on $\gamma_c$) within the lowest order in $t$ \cite{BGM2015} (see also Ref. \cite{Konig2015}). 
For the purpose of this paper, it will be sufficient to use the equations \eqref{eq:rg:final:t} -- \eqref{eq:rg:final:Z} which are of the lowest order in $\gamma_{s,t,c}$  (and are not affected by the above subtleties). 
\color{black}

The renormalization group equations are supplemented by the initial conditions $t=t_0$, $\gamma_{s}=\gamma_{s0}$, $\gamma_{t}=\gamma_{t0}$, and $\gamma_{c}=\gamma_{c0}$ at $y=0$ ($L=\ell$). We assume  that the initial values of resistance and interaction parameters are small, $t_0\ll 1$ and $|\gamma_{s0}|, |\gamma_{t0}|, |\gamma_{c0}|\ll 1$. In addition, we limit our discussion to the case of $t_0\gg |\gamma_{c0}|$. We mention that for $t_0\ll |\gamma_{c0}|$ disorder modifies the BCS transition temperature only slightly. Under the above assumptions, we can neglect the $\gamma_c^2$ term in the r.h.s. of Eq. \eqref{eq:rg:final:gc} at the initial stage of the renormalization group flow. Then, the interaction parameters flow towards the BCS line, 
$$-\gamma_s=\gamma_t=\gamma_c=\gamma.$$ 
The initial value of $\gamma$ is equal to $\gamma_0=(3\gamma_{t0}+2\gamma_{c0}-\gamma_{s0})/6$. 
We assume that $\gamma_0<0$. Projecting renormalization group equations \eqref{eq:rg:final:t} -- \eqref{eq:rg:final:gc} onto the BCS line, 
we find
\begin{equation}
\frac{dt}{dy}=t^2 , \qquad \frac{d\gamma}{dy}=2t \gamma-2\gamma^2/3 \ .
\label{eq:RG:proj}
\end{equation}
{\color{orange}}  Let us emphasize that the disorder-induced correction to  
the renormalization group equation (the term $2t \gamma$ in Eq. \eqref{eq:RG:proj} is intimately related to the multifractality of wave functions in the noninteracting theory. The coefficient $2t$ coincides with the lowest-order term in the expansion of the absolute value of the multifractal exponent $|\Delta_2|$ in powers of the resistance $t$ \cite{Wegner}. As known \cite{Evers}, the exponent $\Delta_2<0$ governs the scaling behavior of the disorder-averaged fourth moment of the wave functions. In the 
NLSM formalism, the scaling of the fourth moment of the wave functions is governed by an operator bilinear in $Q$. Therefore, to the linear order in interaction, the renormalization-group equations of the interacting theory are controlled by scaling dimensions at the noninteracting fixed point \cite{BGM2012,Foster2012}. 
\color{black}

According to  the renormalization group equations \eqref{eq:RG:proj}, $\gamma$ flows towards negative values and diverges eventually at a finite length scale $L_c$. For $|\gamma_0|\gg t_0^2$ one can estimate this lengthscale as  \cite{BGM2012}
\begin{equation}
L_c= \ell \exp\bigl (1/t_0-1/t^{\rm RG}_c\bigr ),
\label{eq:RG:Lc}
\end{equation}
where 
\begin{equation}
t_c^{\rm RG} = 3 t_0^2/(2|\gamma_0|) .
\label{eq:t:RG}
\end{equation}
The divergence of $\gamma$ at the finite length scale signals instability of the normal phase towards superconducting order at temperature $T_c^{\rm RG}=D/L_c^{2}$.
Hence, we obtain 
\begin{equation}
T_c^{\rm RG} \sim \frac{1}{\tau} \exp\left (-\frac{2}{t_0} +\frac{2}{t_c^{\rm RG}}\right ) \ .
\label{eq:Tc:RG}
\end{equation}
We emphasize that $T_c^{\rm RG}$ is enhanced parametrically  in comparison with the clean BCS transition temperature 
$T_c^{\rm BCS}$ given by Eq. \eqref{Tc-BCS}. 
{\color{orange}} This enhancement occurs due to the positive sign of the disorder-induced term $2t\gamma$ in Eq. \eqref{eq:RG:proj}. As explained above, this is a consequence of the multifractality in the noninteracting theory. Therefore, the origin of the enhancement of the superconducting transition temperature is deeply rooted into the multifractality of two-dimensional disordered electrons. \color{black}
Also we note that at $t_0\sim \sqrt{|\gamma_0|}$ transition into the insulating phase occurs \cite{BGM2012}.

\subsection{Relation between renormalization group flow and the modified self-consistency equation\label{sec:RGvsSC}}

We note that Eq. \eqref{eq:gamma:sup} has a striking resemblance with the renormalization group equation for $\gamma_c$ in the normal state, Eq. \eqref{eq:rg:final:gc}. As we shall demonstrate below, this fact is not occasional. As usual, in order to estimate the superconducting transition temperature, we can linearize Eq. \eqref{eq:spectral:gap:2}. Then we find
\begin{equation}
\Delta_\varepsilon= - 2\pi T \sum_{\varepsilon^\prime>0} \gamma_c(L_{\varepsilon+\varepsilon^\prime})  \frac{\Delta_{\varepsilon^\prime}}{\varepsilon^\prime}  \  .
\label{eq:spectral:gap:3}
\end{equation} 

In order to make connection of Eq. \eqref{eq:spectral:gap:3} with the renormalization group equations in the normal state, following Refs. \cite{Skvortsov2005,Feigelman2012}, we substitute  $L_{\varepsilon+\varepsilon^\prime}$ by $\min\{L_{\varepsilon},L_{\varepsilon^\prime}\}$. We also substitute the sum over Matsubara energies by an integral over continuous energies (see below for arguments as to why it is possible). 
Then Eq. \eqref{eq:spectral:gap:3} can be reduced to the following differential equation,
\begin{equation}
\frac{d^2\Upsilon_\varepsilon}{d y_\varepsilon^2} = - t(\gamma_s-3\gamma_t) \bigl [ \Upsilon_\varepsilon - \Upsilon_{\varepsilon=\pi T}\bigr ] \ , \qquad \frac{d^2\Upsilon_\varepsilon}{d y_\varepsilon^2} \Biggl |_{\varepsilon=\pi T}=0 \ , \qquad
\frac{d\ln \Upsilon_{\varepsilon}}{dy_{\varepsilon}}\Biggl |_{\varepsilon=1/\tau}= 2\gamma_{c0} \ , 
\label{eq:Delta_eps:simple}
\end{equation}
where $\Delta_\varepsilon = - (1/2) d\Upsilon_\varepsilon/dy_\varepsilon$ and $y_\varepsilon=\ln L_\varepsilon/\ell$. 
Equation \eqref{eq:Delta_eps:simple} is worthwhile to compare with the equation 
\begin{equation}
\frac{d^2 \Upsilon}{d y^2} = -t (\gamma_s-3\gamma_t) \Upsilon \ , \qquad 
\frac{d\ln \Upsilon}{dy}\Biggl |_{y=0}= 2\gamma_{c0} \ ,
\label{eq:Delta_eps:simple:2}
\end{equation}
that can be obtained from Eq. \eqref{eq:rg:final:gc} upon the substitution $\gamma_c= (d\Upsilon/dy)/(2\Upsilon)$. Equations \eqref{eq:Delta_eps:simple} -- \eqref{eq:Delta_eps:simple:2} become identical if we impose the additional 
condition $\Upsilon_{\varepsilon=\pi T}=0$. Such condition  for $\Upsilon_{\varepsilon}$ in Eq. \eqref{eq:Delta_eps:simple}  is needed in order to have one-to-one correspondence between 
$\Upsilon_{\varepsilon}$ and $\Delta_{\varepsilon}$.
For $\Upsilon$ in Eq. \eqref{eq:Delta_eps:simple:2}, the condition $\Upsilon=0$ appears automatically, since $\gamma_c$ diverges at the length scale $L_c$. 
Therefore, we can identify $\Upsilon_{\varepsilon}$ with $\Upsilon(L_\varepsilon)$: 
\begin{equation}
\Upsilon_{\varepsilon}\equiv\Upsilon(L_\varepsilon) \ .
\end{equation} 
Then the condition $\Upsilon_{\varepsilon=\pi T}=0$ determines 
a certain temperature $T_c$ for which $L_{\pi T_c} = L_c$. Temperature $T_c$ defined in this way is given by Eq. \eqref{eq:Tc:RG}. Thus, the superconducting transition temperature determined from the self-consistency equation coincides with the one found from the renormalization group equations in the normal phase. It is worthwhile to mention that there is a subtle difference between Eq. \eqref{eq:Delta_eps:simple} and Eq. \eqref{eq:Delta_eps:simple:2}. The variables $t$, $\gamma_s$, and $\gamma_t$ in Eq. \eqref{eq:Delta_eps:simple:2} are governed by renormalization group equations \eqref{eq:rg:final:t} -- \eqref{eq:rg:final:gc}, whereas in Eq. \eqref{eq:Delta_eps:simple} they obey Eqs. \eqref{eq:rg:final:t} -- \eqref{eq:rg:final:gt}
and \eqref{eq:gamma:sup}. In the arguments given above we neglected this subtlety.

\subsection{Estimate for $T_c$ from the modified self-consistency equation}

Now we shall compare the prediction for the transition temperature \eqref{eq:Tc:RG} that stems from the renormalization group flow in the normal state with the one obtained from the accurate analysis of the modified self-consistency equation \eqref{eq:spectral:gap:3}. 
In order to analyse this equation, we project the renormalization group equations onto the BCS line $-\gamma_s=\gamma_t=\gamma_c=\gamma$ and neglect interaction corrections to the renormalization of $t$ (see Sec.  \ref{Sec:Tc:RG}). This implies substitution of $\gamma_c$ by $\gamma$ in Eq. \eqref{eq:spectral:gap:3}, where the length-scale dependence of $\gamma$ is governed by the following renormalization group equations:
\begin{equation}
\frac{dt}{dy} = t^2 \ , \qquad \frac{d\gamma}{dy} = 2 \gamma t \  .
\label{eq:RG:proj:1}
\end{equation}
Solving the above equation for $t(L)$ and $\gamma(L)$, we find
\begin{equation}
t(L) = \frac{t_0}{1-t_0\ln L/\ell} , \qquad \gamma(L) = \gamma_0 \frac{t^2(L)}{t_0^2}\ .
\label{eq:sol:RG:reduced}
\end{equation}

\begin{figure}
\centerline{
		(a)\hspace{-0.3cm} \includegraphics[width = 0.47\textwidth]{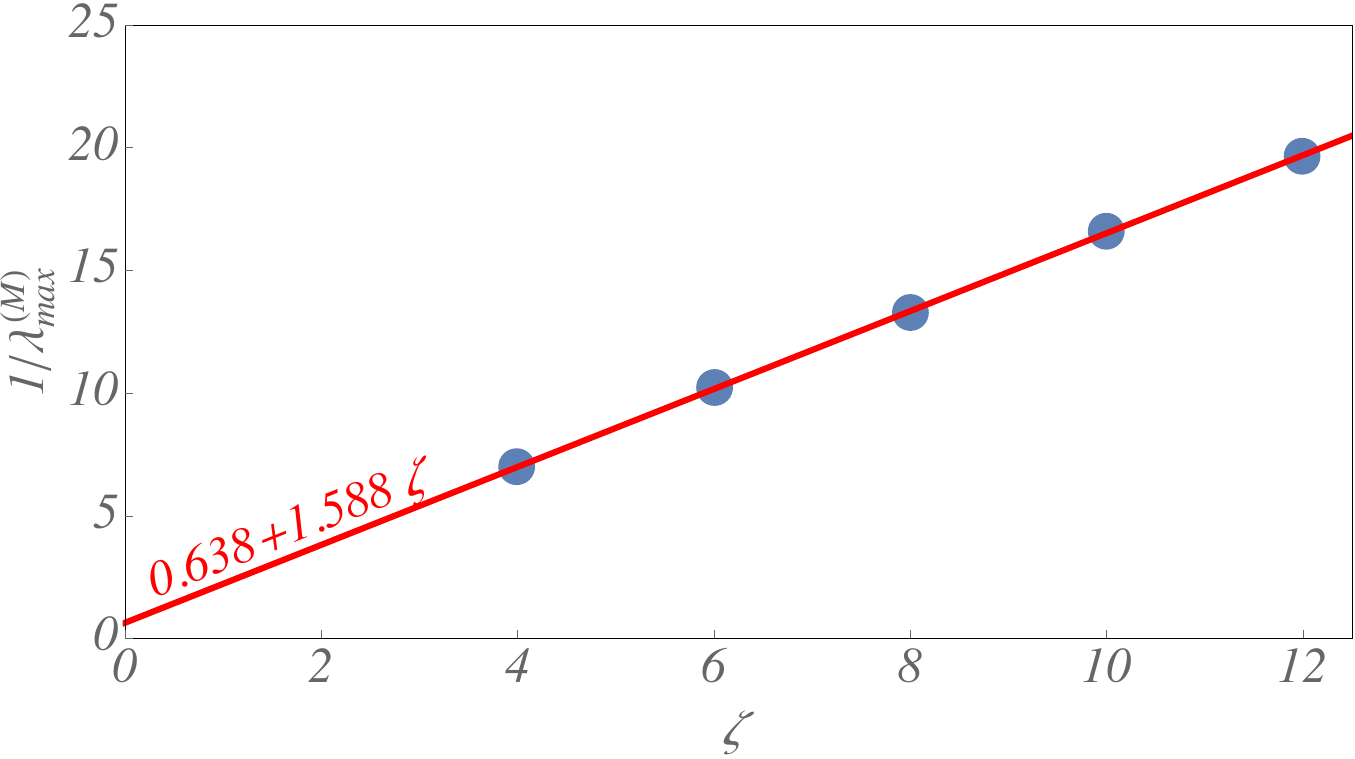} \quad 
		(b)\hspace{-0.3cm} \includegraphics[width = 0.47\textwidth]{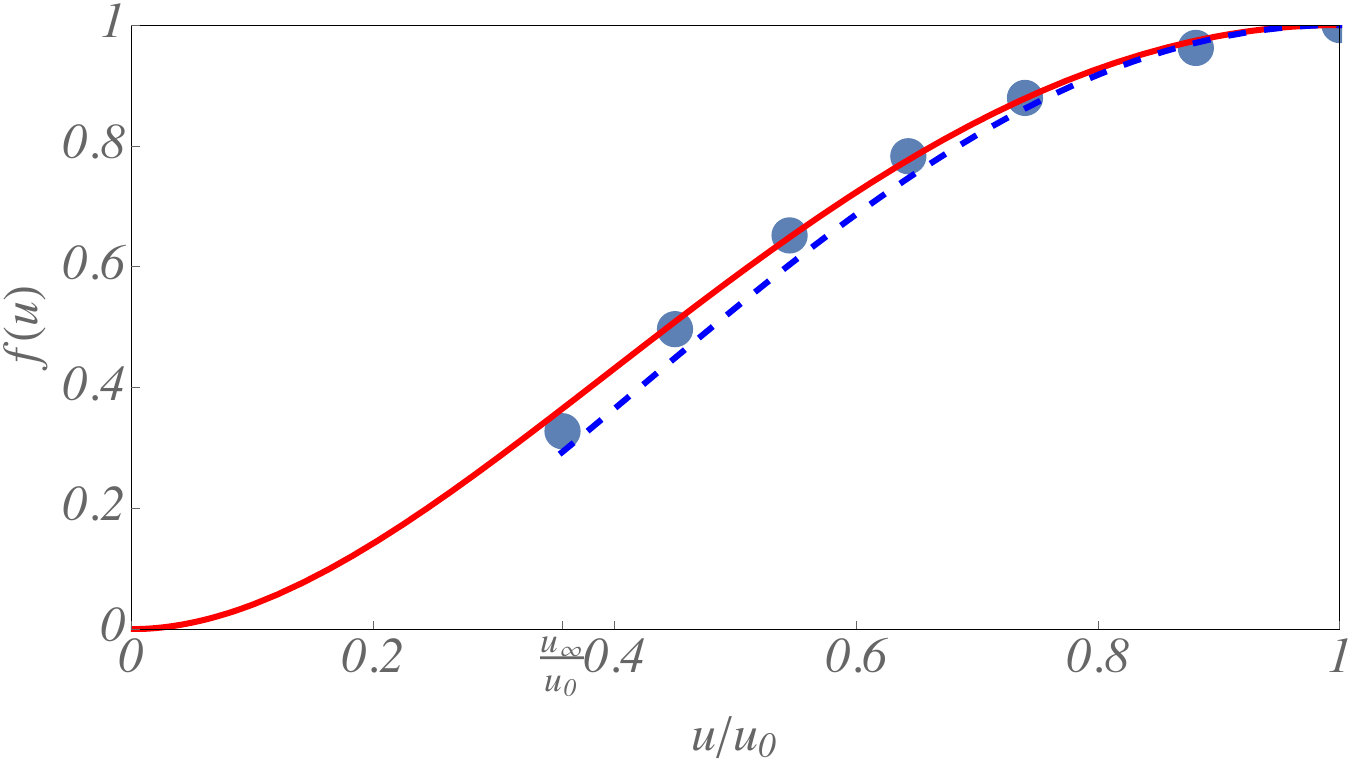}
		}
		\caption{(a) Dependence of the inverse maximum eigenvalue, $1/\lambda_{\max}^{(M)}$ on $\zeta$. The dots are numerical values, the red solid line is a fit. Extrapolation to the infinite size of the matrix $M$ is performed. (b) Comparison between the function $f(z)$, Eq. \eqref{eq:func:F:sol}, (red curve), the function $f(z)$ (blue dot-dashed curve) and the eigenvector (blue dots) corresponding to the maximal eigenvalue of the matrix $M$ for $t_0=0.14$ and $\gamma_0=-0.05$.}
		\label{Fig:1}
\end{figure}

Equation \eqref{eq:spectral:gap:3} can be viewed as an eigenvalue problem, such that the transition temperature is determined by the maximum eigenvalue of a corresponding matrix. We parametrize the transition temperature as $T_c=(2\pi \tau^{-1}) \exp(-2/t_0+2/t_c)$ where $t_c$ satisfies inequality $1\gg t_c\gg t_0$. Then, rewriting Eq. \eqref{eq:spectral:gap:3} with the help of Eq. \eqref{eq:sol:RG:reduced}, we obtain 
\begin{equation}
\Delta_n = \frac{4|\gamma_0|}{t_0^2} \sum_{n^\prime\geqslant 0} 
M_{n,n^\prime}(2/t_c) \Delta_{n^\prime}, \qquad M_{n,n^\prime}(\zeta)=
\frac{1}{[\zeta+\ln(n+n^\prime+1)]^2}\frac{1}{n^\prime+1/2} .
\label{eq:spectral:gap:4}
\end{equation}
The above approximate analysis of the self-consistency equation suggests that in the case $\zeta\gg 1$, the maximum eigenvalue of $M(\zeta)$ behaves as $\lambda_{\max}^{(M)} \propto 1/\zeta$.  The numerical data for $\lambda_{\max}^{(M)}$ can be viewed in Fig. \ref{Fig:1}a.  The numerical results support  the above expectation. For $\zeta\gg 1$, one finds 
$\lambda_{\max}^{(M)} = u_M/(2\zeta)$ with $u_M\approx 1.26$. We note that this result satisfies the 
Perron-Frobenius bound at $\zeta\gg 1$,
\begin{equation}
\lambda_{\max}^{(M)} < \min_{n} \sum_{n^\prime\geqslant 0} M_{n,n^\prime}(\zeta) = 1/\zeta\ . 
\end{equation}
Therefore, the self-consistency equation \eqref{eq:spectral:gap:4} results in the following expression for the superconducting transition temperature, cf. Eq. \eqref{eq:Tc:RG}:
\begin{equation}
T_c^{\rm SC} \sim \frac{1}{\tau} \exp \left (-\frac{2}{t_0}+\frac{2}{t_c^{\rm SC}}\right ) \ , \qquad t_c^{\rm SC} = u_M t_0^2/|\gamma_0| \ .
\label{eq:Tc:SC}
\end{equation}
We note that the assumption $\zeta=2/t_c \gg 1$ is indeed satisfied.

The fact that $\zeta$ is large allows us to find the eigenvector $\Delta_n$ that corresponds to the maximum 
eigenvalue $\lambda_{\max}^{(M)}$. Let us use the Euler-Maclaurin formula for the summation over $n^\prime$ in Eq. \eqref{eq:spectral:gap:4}. This is justified by the condition $\zeta\gg 1$. In addition, we approximate $\ln(n+n^\prime+1)$ by 
$\ln(\max\{n,n^\prime\}+1)$ (as we shall see below, such simplification provides qualitatively correct results). 
With these approximations, we arrive at the following equation:
\begin{equation}
\Delta(u_n)=2 u_n^2 \int \limits_{u_n}^{u_0} \frac{du}{u^2}\ \Delta(u)+ 
2 \int \limits_{u_\infty}^{u_n} du\  \Delta(u) + \frac{t_0^2}{|\gamma_0|} u_n^2 \Delta(u_0)\ ,
\end{equation}
where 
$$u_n = \frac{2 |\gamma_0|}{t_0^2[2/t_c+\ln(n+1)]}.$$ 
The quantity $u_\infty = |\gamma_0|/t_0 \ll 1$ corresponds to $u_n$ with the maximally allowed index 
$n\simeq 1/(2\pi T_c\tau)$.
Let us parametrize $\Delta(u_n)$ as $\Delta(u_n)=\Delta(u_0) f(u_n)$, where a function $f(u)$ satisfies the normalization condition $f(u_0)=1$. Then the function $f(u)$ satisfies the following differential equation, 
\begin{equation}
u f^{\prime\prime}(u) - f^\prime(u) + 4 f(u) = 0 \ , \qquad f^\prime(u_0)=C u_0\ , \qquad f^\prime(u_\infty) = 2 f(u_\infty)/u_\infty \ ,
\label{eq:func:F}
\end{equation}
where $C=2t_0^2/|\gamma_0| \ll 1$. Using the smallness of $u_\infty$, the solution of the above equation can be 
written as
\begin{equation}
f(u)=  \frac{F(u)}{F(u_0)}\ , \qquad F(u) = u J_2(4\sqrt{u}) \  .
\label{eq:func:F:sol}
\end{equation}
Here, $J_2(x)$ denotes the Bessel function of the first kind. The yet unknown parameter $u_0$ can be found from the boundary condition at $u=u_0$, $f^\prime(u_0)=C u_0$. To the lowest order in $C\ll 1$, the parameter $u_0$ can be found as 
\begin{equation}
u_0 = u_c -\frac{C u_c^2}{4} ,
\label{eq:rel:z0:zc}
\end{equation} 
where $u_c\approx 0.92$ is the solution of the equation $F^\prime(u_c)=0$.
Hence, neglecting the difference between $u_0$ and $u_c$ (proportional to $t_0^2/|\gamma_0|\ll 1$), we find that the superconducting transition temperature is given by [cf. Eqs. \eqref{eq:Tc:RG} and \eqref{eq:Tc:SC}]
\begin{equation}
T_c \sim \frac{1}{\tau} \exp \left (-\frac{2}{t_0}+\frac{2}{t_c}\right ) \ , \qquad t_c = u_c t_0^2/|\gamma_0| \ .
\label{eq:Tc:SC:App}
\end{equation}
We note that the 25$\%$ discrepancy in the numerical values of $u_c$ and $u_M$ indicates that the approximation $\ln(n+n^\prime+1)$ by $\ln(\max\{n,n^\prime\}+1)$ is not justified by any small parameter. Nevertheless, our approximate solution for the eigenvector with the maximum eigenvalue, Eq. \eqref{eq:func:F:sol}, is in good agreement with the eigenvector determined numerically, as shown in Fig. \ref{Fig:1}b.

\color{black}
The estimate \eqref{eq:Tc:RG} for the superconducting temperature 
from the analysis of the renormalization group equations in the normal phase and the estimates \eqref{eq:Tc:SC} and \eqref{eq:Tc:SC:App} on the basis of the self-consistency equation are essentially the same except for different numerical values of the constant $u$. In all three cases its value is of the order of unity ($1.5$, $1.26$, and $0.92$, respectively). 
\color{black}
We emphasize also that significant dependence of the gap function $\Delta_\varepsilon$ on the Matsubara energy $\varepsilon$ appears just at the superconducting transition temperature.

\section{The energy dependence of the gap function\label{sec:GF}}

\subsection{The gap function near $T_c$}

The non-trivial energy dependence of the gap function at the transition temperature suggests that $\Delta(z_0)$ vanishes at $T=T_c$. The dependence of the gap function on temperatures at $T_c-T\ll T_c$ can be found in a way similar to the BCS theory. We expand the modified self-consistency equation \eqref{eq:spectral:gap:2}
to the third order in the gap function:
\begin{equation}
\Delta_\varepsilon = 2\pi T \sum_{\varepsilon^\prime> 0} |\gamma(L_{\varepsilon+\varepsilon^\prime})| \frac{\Delta_{\varepsilon^\prime}}{\varepsilon^\prime} - \pi T \sum_{\varepsilon^\prime> 0} |\gamma(L_{\varepsilon+\varepsilon^\prime})| \frac{\Delta_{\varepsilon^\prime}^3}{\varepsilon^{\prime 3}} \ .
\label{eq:parabolic}
\end{equation}
We note that the expansion of $\gamma\left (
L_{\sqrt{\varepsilon^{2}+\Delta_{\varepsilon}^2}+\sqrt{\varepsilon^{\prime 2}+\Delta_{\varepsilon^\prime}^2}}
\right )$ in powers of $\Delta_\varepsilon$ and $\Delta_{\varepsilon^\prime}$ leads to the terms which are proportional to a small factor $t_0^4/\gamma_0^2 \ll 1$. 
Such terms can be safely neglected.
Let us parameterize the gap function as $\Delta_\varepsilon = \Delta_0(T) f(u_\varepsilon)$ where $u_\varepsilon = |\gamma_0|t(L_\varepsilon)/t_0^2$. Then taking into account that the sum over $\varepsilon^\prime$ in the last term of the right-hand side of Eq. \eqref{eq:parabolic} is dominated by $\varepsilon^\prime\sim T_c$, we find
\begin{gather}
f(u_\varepsilon) = 2 u_\varepsilon^2 \int \limits_{u_\varepsilon}^{u_T} \frac{du}{u^2} f(u) + 2 \int \limits_{u_\infty}^{u_\varepsilon} du f(u) + \frac{C}{2} u_\varepsilon^2, \qquad 
C = \frac{2 t_0^2}{|\gamma_0|} \left (1 - \frac{7\zeta(3) \Delta_0^2(T)}{8\pi^2 T_c^2}  f^3(z_T)\right ) \ ,
\end{gather}
where $u_T = |\gamma_0| t(L_{\pi T})/t_0^2$. The above integral equation is reduced to Eq. \eqref{eq:func:F} but with $u_T$ instead of $u_0$. Using Eq. \eqref{eq:rel:z0:zc} and expressing $u_T$ and $u_0$ in terms of $T$ and $T_c$, respectively, we obtain the following well-known result of the BCS theory:
\begin{equation}
\Delta_0(T)= \left [ \frac{8\pi^2}{7\zeta(3)}T_c(T_c-T)\right ]^{1/2} \ , \qquad T_c-T\ll T_c \  .
\end{equation}
We note that the corrections to the BCS-type temperature dependence are controlled by the small parameter $t_0^2/|\gamma_0| \ll 1$.

\subsection{The gap function at low temperatures $T\ll T_c$}

Let us first analyze the energy dependence of the gap function $\Delta_\varepsilon$ at the zero temperature. We expect that the function $\Delta_\varepsilon$ has a form similar to the one at $T=T_c$, see Fig. \ref{Fig:1}b.
We introduce the energy $\varepsilon_0$ which is given by the solution of the equation $\varepsilon_0 = \Delta_{\varepsilon_0}$. Then, we assume that for $\varepsilon<\varepsilon_0$ the gap function $\Delta_\varepsilon$ is close to its value $\Delta_0$ at $\varepsilon=0$. For $\varepsilon>\varepsilon_0$ the gap function $\Delta_\varepsilon$
is a monotonously decreasing function that reaches the value $\Delta$ at $\varepsilon \sim 1/\tau$. 

At $\varepsilon<\varepsilon_0$ the self-consistency equation \eqref{eq:spectral:gap:2}
can be approximately written as 
\begin{align}
\Delta_\varepsilon  = & 
\int \limits_{0}^{1/\tau} \frac{d\varepsilon^\prime \Delta_{\varepsilon^\prime}}{\sqrt{\varepsilon^{\prime 2}+\Delta_{\varepsilon^\prime}^2}} 
\bigl |\gamma(L_{\Delta_0+\sqrt{\varepsilon^{\prime 2}+\Delta_{\varepsilon^\prime}^2}})\bigr | \left [1-\frac{\sqrt{\varepsilon^2+\Delta_{\varepsilon}^2}-\Delta_0}{\Delta_0+\sqrt{\varepsilon^{\prime 2}+\Delta_{\varepsilon^\prime}^2}} t(L_{\Delta_0+\sqrt{\varepsilon^{\prime 2}+\Delta_{\varepsilon^\prime}^2}})\right ]
\notag \\
= & \Delta_0 - \bigl [\sqrt{\varepsilon^2+\Delta_{\varepsilon}^2}-\Delta_0\bigr ]
\int \limits_{0}^{1/\tau} \frac{d\varepsilon^\prime \Delta_{\varepsilon^\prime}}{\sqrt{\varepsilon^{\prime 2}+\Delta_{\varepsilon^\prime}^2}} \frac{
\bigl |\gamma(L_{\Delta_0+\sqrt{\varepsilon^{\prime 2}+\Delta_{\varepsilon^\prime}^2}})\bigr |
t(L_{\Delta_0+\sqrt{\varepsilon^{\prime 2}+\Delta_{\varepsilon^\prime}^2}})}{\Delta_0+\sqrt{\varepsilon^{\prime 2}+\Delta_{\varepsilon^\prime}^2}}
 \ . 
\label{eq:gap:1}
\end{align}
Since the dependence of $\gamma$ on $L$, governed by Eq. \eqref{eq:sol:RG:reduced}, is only logarithmical, the integral over $\varepsilon^\prime$ in Eq. \eqref{eq:gap:1} is dominated by $\varepsilon^\prime \sim \varepsilon_0$. We then find:
\begin{equation}
\Delta_\varepsilon  \simeq 
\Delta_0 - \frac{t_0^4}{\gamma_0^2} u^3_{\Delta_0+\sqrt{2}\varepsilon_0} \left [\sqrt{\varepsilon^2+\Delta_{\varepsilon}^2}-\Delta_0\right ] \int\limits_0^\infty \frac{dx}{\sqrt{1+x^2}(\Delta_0/\varepsilon_0+\sqrt{1+x^2})}\  .
\label{eq:GapFunction:small:0}
\end{equation}
This result implies that $\Delta_\varepsilon$ is constant at $\varepsilon<\varepsilon_0$ up to corrections of the order of $t_0^4/\gamma_0^2$. With the same accuracy the energy scale $\varepsilon_0$ coincides with $\Delta_0$. Further, we can substitute $u_{\Delta_0}$ for $u_{\Delta_0+\sqrt{2}\varepsilon_0}$ in Eq. \eqref{eq:GapFunction:small:0}. Therefore, at $\varepsilon<\Delta_0$ the gap function behaves as 
\begin{equation}
\Delta_\varepsilon  \simeq 
\Delta_0 - \frac{t_0^4}{\gamma_0^2} u^3_{\Delta_0} \left [\sqrt{\varepsilon^2+\Delta_{0}^2}-\Delta_0\right ] \  .
\label{eq:GapFunction:small}
\end{equation}

At $\varepsilon\geqslant \varepsilon_0\simeq\Delta_0$ we approximate the self-consistency equation \eqref{eq:spectral:gap:2} as follows:
\begin{equation}
\Delta_\varepsilon = 
\bigl |\gamma(L_{\varepsilon})\bigr |
\int \limits_0^{\varepsilon_0} \frac{d\varepsilon^\prime \Delta_{0}}{\sqrt{\varepsilon^{\prime 2}+\Delta_{0}^2}} 
+\bigl |\gamma(L_{\varepsilon})\bigr |
\int \limits_{\varepsilon_0}^{\varepsilon} \frac{d\varepsilon^\prime \Delta_{\varepsilon^\prime}}{\varepsilon^{\prime}} 
+\int \limits_{\varepsilon}^{1/\tau} \frac{d\varepsilon^\prime \Delta_{\varepsilon^\prime}}{\varepsilon^{\prime}} 
\bigl |\gamma(L_{\varepsilon^\prime})\bigr |  \  .
\label{eq:gap:2}
\end{equation}
Here, the dependence of $\gamma$ on $L$ is governed by Eq. \eqref{eq:sol:RG:reduced}. We note that Eq. \eqref{eq:gap:2} is justified for $\varepsilon \gg \varepsilon_0$. For $\varepsilon\simeq \varepsilon_0$, a
more accurate treatment of the self-consistency equation \eqref{eq:spectral:gap:2} 
results in corrections of the order of $t_0^2/|\gamma_0|$. Parametrizing the energy dependence of the gap function at $\varepsilon\geqslant \varepsilon_0$ as $\Delta_\varepsilon = \varepsilon_0 f(u_\varepsilon)$, we find that the integral equation \eqref{eq:gap:2} results in Eq. \eqref{eq:func:F} with  $u_{\varepsilon_0}$ entering in place of $u_{0}$ and with the value of the constant $C=2 c_1 t_0^2/|\gamma_0|$, where $c_1=\arcsinh(1)$. We note that, in order to find the precise value of $c_1$, one needs to take into account the $t_0^2/|\gamma_0|$ corrections to Eq. \eqref{eq:gap:2}. However, we are not interested in such accuracy and set $C$ to zero.  Then, we find that the function $f(u)$ is given by Eq. \eqref{eq:func:F:sol} with $u_c$ replacing $u_0$. With the same accuracy, we have $\varepsilon_0=\Delta_0 \simeq T_c$, cf. Eq. \eqref{eq:Tc:SC:App}. 

\begin{figure}
		\centering
		\includegraphics[width = 0.5\textwidth]{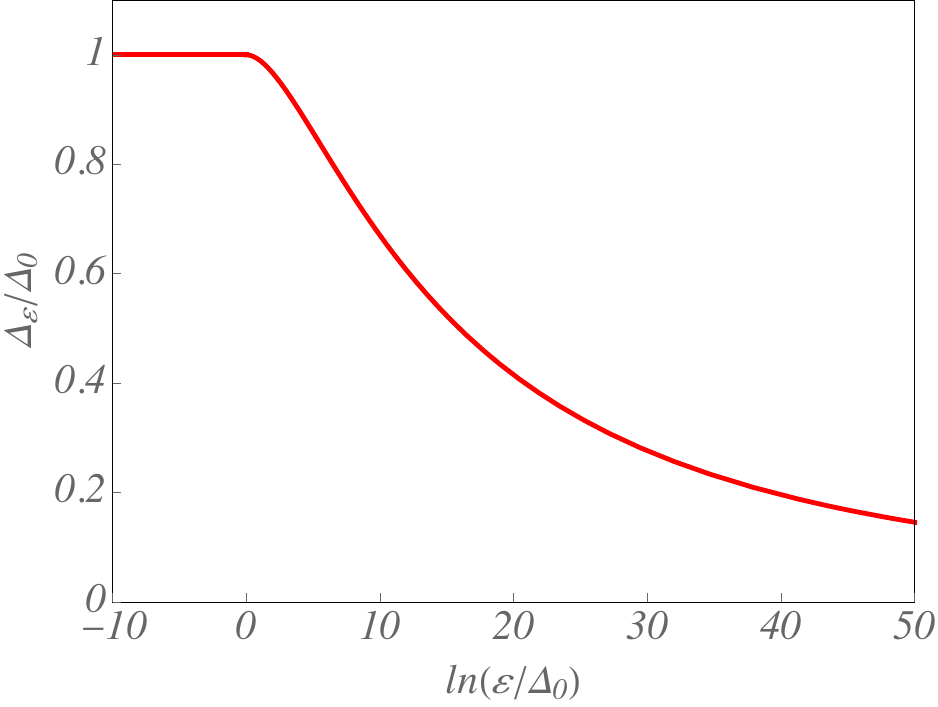} 
		\caption{Dependence of ratio $\Delta_\varepsilon/\Delta_0$ on $\ln( \varepsilon/\Delta_0)$ at $T=0$ for $\gamma_0=-0.005$ and $t_0=0.03$. The monotonously decreasing part of the curve corresponds to the function $f(z_\varepsilon)$.}
		\label{Fig:Delta}
\end{figure}

Expressing the right-hand side of Eq. \eqref{eq:gap:2} in terms of the function $f(z)$ and setting $\varepsilon\sim 1/\tau$, we retrieve 
\begin{equation}
\Delta = c_\Delta (\gamma_0/t_0)^2 \Delta_0 \ , \qquad 
c_\Delta = 
 2 \int \limits_{0}^{u_{c}} \frac{du}{u^2} f(u)  \approx 5.4 \ .
\label{eq:Del:1}
\end{equation}
We note that $\Delta \ll \Delta_0$. All in all, we can summarize the energy dependence of the spectral gap at $T=0$ as 
\begin{equation} 
\Delta_\varepsilon = \Delta_0 
\begin{cases} 
1  & , \quad \varepsilon <\Delta_0\ , \\
\displaystyle \frac{F\bigl(|\gamma_0|t(L_\varepsilon)/t_0^2\bigr)}{F(u_c)}
& , \quad \varepsilon \geqslant\Delta_0 \ ,
\end{cases}
\qquad\qquad  \Delta_0  \sim \frac{1}{\tau} \exp \left (-\frac{2}{t_0}+\frac{2 |\gamma_0|}{u_c t_0^2}\right ) \ .
\label{eq:gap:function:final:1}
\end{equation} 
The overall dependence of the gap function $\Delta_\varepsilon$ on $\varepsilon$ is shown in Fig. \ref{Fig:Delta}.

We emphasize that within our approximate treatment, the values of $\Delta_0$ and $T_c$ coincide with the exponential accuracy, \color{black} i.e. the exponential factor in Eq. \eqref{eq:gap:function:final:1} coincides with the exponential factor in Eq. \eqref{eq:Tc:SC:App}. However, we cannot exclude a possibility that the magnitude of the ratio $\Delta_0/T_c$ differs from the one in the BCS theory. This can result from the  
corrections of the order of $t_0^2/|\gamma_0|$ that determine the coefficient $C$ in Eq. \eqref{eq:func:F}, see Eq. \eqref{eq:rel:z0:zc}. \color{black}

In order to consider the effect of nonzero but low temperatures, $T\ll T_c$, we need to perform summation over Matsubara frequencies in the interval $0<\varepsilon_n \lesssim \Delta_0$ more accurately. We assume that the parameter $\Delta_0$ becomes temperature dependent, $\Delta_0 \to \Delta_0(T)$. Then Eq. \eqref{eq:gap:2} should be modified as follows:
\begin{gather}
\Delta_\varepsilon  = 
 |\gamma(L_{\varepsilon})|
 \left ( 2\pi T \sum_{\varepsilon^\prime>0} \frac{\Delta_0(T)}{\varepsilon^{\prime 2}+\Delta_0^2(T)} - \int \limits_{\Delta_0(T)}^{1/\tau}   \frac{d\varepsilon^\prime \ \Delta_0(T)}{\sqrt{\varepsilon^{\prime 2}+\Delta_0^2(T)}} \right )+\int \limits_{\varepsilon_0}^{\varepsilon} \frac{d\varepsilon^\prime}{\varepsilon^\prime} |\gamma(L_{\varepsilon})| \Delta_{\varepsilon^\prime}
 \notag \\
+ \int\limits_{\varepsilon}^{1/\tau} \frac{d\varepsilon^\prime}{\varepsilon^\prime} |\gamma(L_{\varepsilon^\prime})| \Delta_{\varepsilon^\prime} \ .
\label{eq:Delta:TllTc}
\end{gather}
Using the parametrization, $\Delta_\varepsilon=\Delta_0(T) f(u_\varepsilon)$, one can check that the function $f(u)$ satisfies Eq. \eqref{eq:func:F} with $u_0$ replaced by $u_0(T) = |\gamma_0| t(L_{\Delta_0(T)})/t_0^2$. The constant $C$ becomes a $T$-dependent function given by 
\begin{equation}
C(T) \simeq C(T=0)-
\frac{4t_0^2}{|\gamma_0|} \int \limits_0^\infty \frac{dE}{\sqrt{E^2+\Delta_0^2}} e^{- \sqrt{E^2+\Delta_0^2}/T} 
= C(T=0)- \frac{2t_0^2}{|\gamma_0|} \sqrt{\frac{2\pi T}{\Delta_0}} e^{-\Delta_0/T} \ .
\end{equation}
Using Eq. \eqref{eq:rel:z0:zc}, we can express $u_0(T)$ in terms of $u_0$, yielding 
\begin{equation}
\Delta_0(T)  = \Delta_0 -  (2\pi T \Delta_0)^{1/2} e^{-\Delta_0/T}\ , \qquad T\ll T_c \ .
\end{equation}
Therefore, the dependence of the magnitude of the gap function on temperature $T\ll T_c$ at $\varepsilon< \Delta_0$ 
is the same as in the BCS theory. We note that there are corrections to this result which are of the order of $t_0^2/|\gamma_0|$.

\section{Disorder-averaged density of states \label{Sec:AvDOS}}

In this section, we analyze the disorder-averaged density of states of single-particle excitations in the superconducting state.
Within the NLSM formalism, the average density of states can be obtained as
\begin{equation}
\rho(E) = \frac{\nu}{4} \re \spp \langle Q_{nn}^{\alpha\alpha} \rangle \ ,
\label{eq:dos:b1}
\end{equation}
where the analytic continuation to real energies, $\varepsilon_n\to E+i 0$, is performed.
Here symbol `$\spp$' denotes the trace over particle-hole and spin spaces. Plugging in the parametrization  \eqref{eq:Q-W} into Eq. \eqref{eq:dos:b1}, one finds
\begin{equation}
\rho(E)= \nu \re Z^{1/2}_\varepsilon \cos\theta_\varepsilon \ \Biggl |_{\varepsilon\to -i E+0} =  \nu \re Z^{1/2}_\varepsilon \frac{\varepsilon}{\sqrt{\varepsilon^2+\Delta^2_{\varepsilon}}} \ \Biggl |_{\varepsilon\to -i E+0} \ .
\label{eq:DOS:def}
\end{equation}
The factor $Z_\varepsilon$ can be written as $Z_\varepsilon = Z(L_{\sqrt{\varepsilon^2+\Delta_\varepsilon^2}})$, where \cite{BGM2016}
\begin{equation}
\frac{d \ln Z}{d\ln y} = t (\gamma_s+3\gamma_t+\gamma_c) .
\label{eq:RG:Z}
\end{equation}
Projecting Eq. \eqref{eq:RG:Z} to the BCS line, one finds that 
\begin{equation}
Z_\varepsilon \simeq 1 + 2 \gamma(L_{\sqrt{\varepsilon^2+\Delta_\varepsilon^2}}).
\end{equation}
This factor encodeds the interaction-induced corrections to the density of states of the type that leads to the zero-bias anomaly
in the normal state \cite{BGM2016}. 

As one can see, in order to compute the average density of states, we need to make an analytic continuation for the gap function $\Delta_\varepsilon$ from Matsubara energies to real energies. We emphasize that the nontrivial dependence of $\Delta_\varepsilon$ on $\varepsilon$ implies that the spectral gap at real energies, $\Delta(E)$, has both real and imaginary parts. The symmetry implies that  $\re \Delta(E)$ is an even function of $E$ whereas $\im \Delta(E)$ is an odd function of $E$.

\subsection{The real and imaginary parts of $\Delta(E)$}

We restrict our consideration to the case of zero temperature, $T=0$. Performing analytic continuation in Eq. \eqref{eq:GapFunction:small}, we find that the gap function at $E\leqslant \Delta_0$ is purely real,
\begin{equation}
\Delta(E)  \simeq 
\Delta_0 - \frac{t_0^4 u^3_{0}}{\gamma_0^2} \left [\sqrt{\Delta_{0}^2-E^2}-\Delta_0\right ] \  .
\label{eq:GapFunction:small:real}
\end{equation}
At energies $E\gtrsim \Delta_0$, the imaginary part of $\Delta(E)$ appears. To show that the nonzero imaginary part does exist, one can consider the perturbative expression \eqref{eq:spectral:gap}. Performing analytic continuation to the real energy, we find that $\im \Delta(E\sim 1/\tau)\sim \Delta t_0$. We note that the imaginary part of $\Delta(E)$ is nonzero due to the imaginary part of the renormalized interaction in the Cooper channel, $\gamma_c$, which corresponds to the superconducting fluctuation propagator. 

For $\varepsilon\gtrsim \Delta_0$, we 
energy dependence of the gap function is described by Eq. \eqref{eq:gap:function:final:1}. After analytic continuation, $i\varepsilon\to E+i 0$, in Eq. \eqref{eq:gap:function:final:1}, we obtain
\begin{equation}
\Delta(E) = \Delta_0 f(\tilde{u}_E) \ ,
\end{equation}
where
\begin{equation}
\tilde{u}_E=\frac{|\gamma_0| \tilde{t}(L_E)}{t_0^2} = \frac{|\gamma_0|/t_0}{1+(t_0/2) \ln (E\tau)-i \pi t_0/4} \simeq 
u_E \left ( 1 + \frac{i\pi t_0^2}{4|\gamma_0|} u_E\right ) \ .
\end{equation}
Here, we have used the fact that $t(L_E)\ll 1$. 
The real and imaginary parts of $\Delta(E)$ read:
\begin{subequations}
\begin{align}
\re \Delta(E) = & \Delta_0 f\left (\frac{|\gamma_0| t(L_E)}{t_0^2}\right) \ , 
\label{eq:Delta:E:Large:Re}
\\
\im \Delta(E) = & \Delta_0 \frac{\pi |\gamma_0| t^2(L_E)}{4 t_0^2} f^\prime \left (
\frac{|\gamma_0| t(L_E)}{t_0^2}\right )
 \ . 
 \label{eq:Delta:E:Large:Im}
\end{align}
\end{subequations}

Expression \eqref{eq:Delta:E:Large:Re} has the same range of validity as Eq. \eqref{eq:gap:function:final:1}. It provides us with an estimate for the real part of the gap function at $E\geqslant \Delta_0$ withcorrections of the order of $t_0^2/|\gamma_0|$ neglected. The situation with the imaginary part of the gap function is  more delicate. 
Equation \eqref{eq:Delta:E:Large:Im} means that $\im \Delta(E\sim \Delta_0) \sim \Delta_0 t_0^4/\gamma_0^2$. We note that the contribution of the same order will be given by the continuation of Eq. \eqref{eq:GapFunction:small:real} to energies larger than $\Delta_0$. This fact implies that, in order to compute $\im \Delta(E)$ at energies $E\sim \Delta_0$, one needs to retain terms of the order of $t_0^4/\gamma_0^2$ in the solution of Eq. \eqref{eq:spectral:gap:2}. However, at large energies, $E\gg \Delta_0$, Eq. \eqref{eq:Delta:E:Large:Im} provides the leading result to the imaginary part of the gap function. For example, at $E\sim 1/\tau$, Eq. \eqref{eq:Delta:E:Large:Im} yields $\im \Delta(E\sim 1/\tau) \sim \Delta t_0$ that matches with the perturbative result. Additional argument that the more accurate (to the order $t_0^4/\gamma_0^2$) expression for the real part of $\Delta(E)$  is needed in order to determine $\im \Delta(E)$ is given by the Kramers-Kronig relations for imaginary and real parts of $\Delta(E)$ (see \ref{App:KK}).

The above results suggest that there is an energy $E_{\rm g}$ such that $\im \Delta(E)=0$ for $E\leqslant E_{\rm g}$ and $\im \Delta(E)>0$ for $E>E_{\rm g}$. Making analytic continuation to the real frequencies in the self-consistency equation \eqref{eq:spectral:gap:2}, we obtain
\begin{equation}
\Delta(E) = 2\pi T\sum_{\varepsilon^\prime>0} \bigl |\gamma(L_{\sqrt{\Delta^2(E)-E^2}+\sqrt{\varepsilon^{\prime 2}+\Delta_{\varepsilon^\prime}^2}})\bigr |\frac{\Delta_{\varepsilon^\prime}}{\sqrt{\varepsilon^{\prime 2}+\Delta_{\varepsilon^\prime}^2}} \ .
\label{eq:DeltaE:1}
\end{equation} 
For energies close to the energy $E_{\rm g}$, we can expand the right hand side of Eq. \eqref{eq:DeltaE:1} as
\begin{subequations}
\begin{gather}
\Delta(E)  = \Delta(E_{\rm g}) - \alpha 
 \Biggl [\sqrt{\Delta^2(E)-E^2} - \sqrt{\Delta^2(E_{\rm g})-E_{\rm g}^2}\ \Biggr ] \ ,
 \label{eq:DeltaE:2a} \\
 \alpha  = 2\pi T\sum_{\varepsilon^\prime>0} \bigl |\gamma(L_{\sqrt{\Delta^2(E_{\rm g})-E_{\rm g}^2}+\sqrt{\varepsilon^{\prime 2}+\Delta_{\varepsilon^\prime}^2}})\bigr | \ t(L_{\sqrt{\Delta^2(E_{\rm g})-E_{\rm g}^2}+\sqrt{\varepsilon^{\prime 2}+\Delta_{\varepsilon^\prime}^2}})\frac{\Delta_{\varepsilon^\prime}}{\varepsilon^{\prime 2}+\Delta_{\varepsilon^\prime}^2} \ .
 \label{eq:DeltaE:2b}
\end{gather}
\end{subequations}
We shall demonstrate below that the parameter $\alpha \sim t_0^4/\gamma_0^2\ll 1$. Solving Eq. \eqref{eq:DeltaE:2a} for $\alpha\ll 1$, we find the dependence of the gap function for real energies close to $E_{\rm g}$,
\begin{equation}
\Delta(E) = \Delta(E_{\rm g}) -\alpha \sqrt{E_{\rm g}^2-E^2}\ , \qquad \Delta(E_{\rm g}) = E_{\rm g} (1+\alpha^2/2) \ , \qquad |E-E_{\rm g}|\ll E_{\rm g} \ . 
\label{eq:DeltaE:3}
\end{equation} 
In agreement with our assumptions, $\im \Delta(E)=0$ for $E\leqslant E_{\rm g}$. 
We note that $\Delta(E)\geqslant E$ for $E\leqslant E_{\rm g}$. 
 For $E>E_{\rm g}$ the imaginary part of the gap function is non-zero, $\im \Delta(E)= -i \alpha \sqrt{E^2-E_{\rm g}^2}$. We note that negative sign of $\im \Delta(E)$ at energies $E>E_{\rm g}$ is needed for positivity of the density of states (see below). However, away from $E_g$ the imaginary part of $\im \Delta(E)$ has to change sign in order to be consistent with the asymptotic at large energies, Eq. \eqref{eq:Delta:E:Large:Im}. The change of the sign of $\im \Delta(E)$ occurs at the energy of the order of a few $E_g$. 
 
Now we can estimate the parameter $\alpha$. The comparison of Eq. \eqref{eq:DeltaE:3} with Eq. \eqref{eq:GapFunction:small:real} implies that $E_{\rm g}$ is of the order of $\Delta_0$. The sum over $\varepsilon^\prime$ in Eq. \eqref{eq:DeltaE:2b} is dominated by $\varepsilon^\prime \simeq \Delta_{\varepsilon^\prime} \simeq \Delta_0$. Then we find
\begin{equation}
\alpha \simeq \bigl |\gamma(L_{2\Delta_0})\bigr | t(L_{2\Delta_0}) 
\int\limits_0^\infty \frac{d\varepsilon^\prime \ \Delta_0}{\varepsilon^{\prime 2}+ \ \Delta_0^2} = \frac{\pi u_0^3}{2} \frac{t_0^4}{\gamma_0^2} \ .
\end{equation}
Here, we have taken into account that $\sqrt{\Delta^2(E_{\rm g})-E_{\rm g}^2} =\alpha E_{\rm g} \ll \Delta_0$.

\subsection{The average density of states\label{Sec:AvDOS:ss}}

With the help of the above results for $\Delta(E)$, we can derive the expression for disorder--averaged density of states as a function of energy. For large energies, $E\gg \Delta_0$, the imaginary part of $\Delta(E)$ can be neglected in comparison with its real part, see Eqs. \eqref{eq:Delta:E:Large:Re} and \eqref{eq:Delta:E:Large:Im}. Since for $E\gg \Delta_0$ the energy is always larger than the real part of $\Delta(E)$, we find 
\begin{equation}
\rho(E)= \nu \frac{E}{\sqrt{E^2- \Delta_0^2 f^2\left (|\gamma_0| t(L_E)/t_0^2\right )}}  \ ,  \qquad \Delta_0 
\ll E\lesssim  1/\tau \ .
\label{eq:avDOS:final}
\end{equation}
Here and hereinafter we neglect the factor $Z_\varepsilon$, see Eq. \eqref{eq:DOS:def}, since it provides negligible corrections of the order of $t_0^2/|\gamma_0|$ in the parametric regime that we consider.

In order to compute $\rho(E)$ at energies close to $E_{\rm g}\simeq \Delta_0$, we use the expression \eqref{eq:DeltaE:3} for $\Delta(E)$. Introducing $\Delta E= E-E_{\rm g}$, we find for $|\Delta E|\ll E_{\rm g}$, 
\begin{equation}
\rho(E)  
= \frac{\nu}{\alpha} \begin{cases}
0 , & \qquad \Delta E<0 \ ,\\
\im \left [ 1 - \frac{2\Delta E}{\alpha^2E_{\rm g}} - 2i \sqrt{\frac{2\Delta E}{\alpha^2E_{\rm g}}}\right ]^{-\frac{1}{2}} , & \qquad \Delta E\geqslant 0 \ .
\end{cases}
\label{eq:DOS:1}
\end{equation}
As expected, the energy $E_{\rm g}$ determines the gap edge in the disorder-averaged density of states. Using Eq. \eqref{eq:DOS:1}, we derive a square-root dependence of the density of state near the gap edge:
\begin{equation}
\rho(E)  
= {\color{orange}}\nu\color{black} \frac{\sqrt{2}}{\alpha^2} \sqrt{\frac{E-E_{\rm g}}{E_{\rm g}}}\ , \qquad 0<E-E_{\rm g} \ll \alpha^2 E_{\rm g} \ .
\label{eq:DOS:2}
\end{equation}
The square-root growth of $\rho(E)$ turns into  the maximum at $E=E_{\rm max}=E_{\rm g}(1+\alpha^2/2)$. The magnitude at the maximum is 
given as $\rho(E_{\rm max})=\nu/(2\alpha)\sim \nu \gamma_0^2/t_0^4 \gg \nu$. For energies larger than $E_{\rm max}$, the density of states decays,
\begin{equation}
\rho(E)  
= \frac{\nu}{\sqrt{2}} \sqrt{\frac{E_{\rm g}}{E-E_{\rm g}}} \ , \qquad  \alpha^2 E_{\rm g} \ll E-E_{\rm g}  \ll {\color{orange}} E_g \color{black} \ .
\label{eq:DOS:3}
\end{equation}
We note that Eq. \eqref{eq:DOS:3} matches the high energy asymptotics, Eq. \eqref{eq:avDOS:final}, at energy of the order of $E_g$. 

\begin{figure}
		\centering
		\includegraphics[width = 0.5\textwidth]{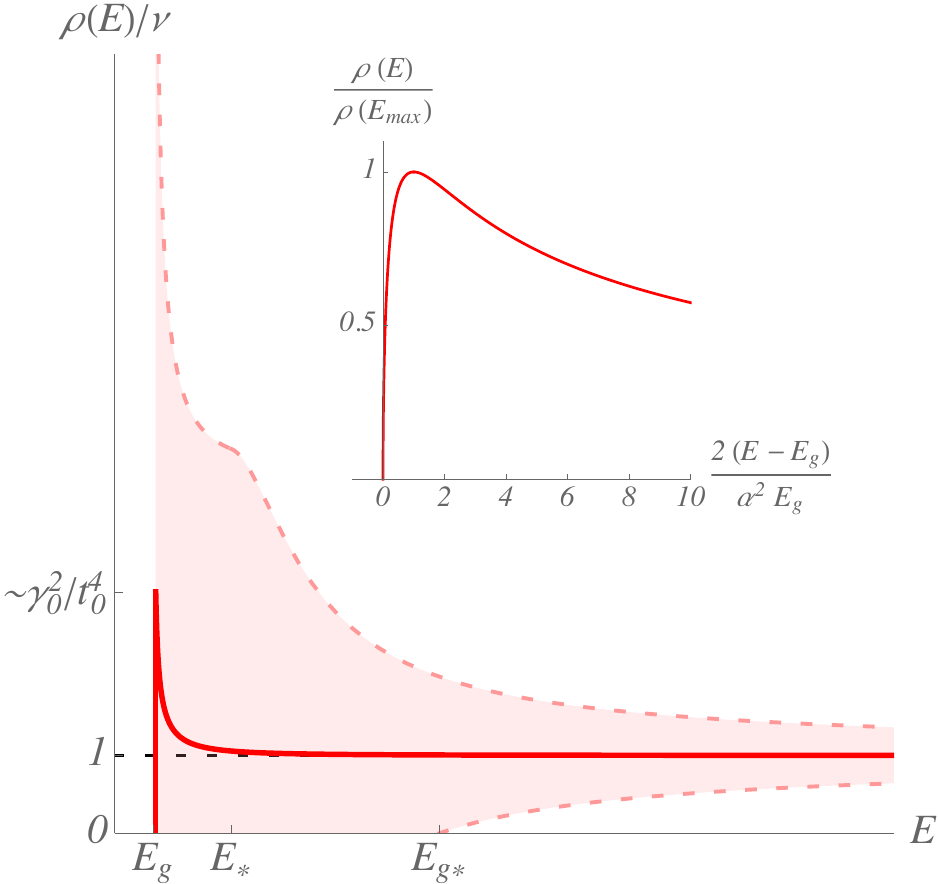} 
		\caption{Sketch of the dependence of the local density of states $\rho(E,\bm{r})/\nu$ on the energy $E$ for $|\gamma_0|\ll t_0\ll \sqrt{|\gamma_0|}$. The red solid curve corresponds to the disorder-averaged density of states (see text in Sec. \ref{Sec:AvDOS:ss}). The dashed red curves and the shaded region illustrates the mesoscopic fluctuations of the local density of states. There are no quasiparticle states below $E_g$. The energy $E_{g*}$ separates the region of strong and weak mesoscopic fluctuations. The energy $E_{*}$ demarcates the energy regions where the mesoscopic fluctuations are cut off by the length scale $L_{*}$ ($E_g<E<E_*$) and the dephasing length $L_{\phi}$ ($E_*<E$) (see text in Sec. \ref{sec:MesoLDOSL:RG}).  Inset: the region close to the gap edge $E_g$. The curve corresponds to Eq. \eqref{eq:DOS:1}.  \color{black}}
		\label{Fig:DOS}
\end{figure}

The dependence of the disorder-averaged density of states on energy is shown in Fig. \ref{Fig:DOS}.  (See Sec. \ref{sec:MesoLDOSL:RG} for discussion of the mesoscopic fluctuations of the local density of states.) \color{black} We note that the profile 
of $\rho(E)$ is very similar to the density of states in the presence of the depairing term in the Usadel equation, introduced, e.g., by the scattering off magnetic impurities \cite{AG1960}. The comparison of Eq. \eqref{eq:DOS:2}, as well as the position of the maximum and its magnitude, with the Abrikosov-Gor'kov theory suggests that the effective depairing parameter $\eta \sim \alpha^{3}$. We note that the depairing term in the Usadel equation appears also in the model with a spatially varying, random interaction in the Cooper channel \cite{LO1972} and with a spatially varying random order parameter \cite{SF2013}.

\section{Mesoscopic fluctuations of the local density of states in the superconducting state \label{sec:MesoLDOS}}

\subsection{Perturbative approach \label{subsec:MesoPert}}

We start analysis of the mesoscopic fluctuations of the local density of states from the calculation of its dispersion. It can be expressed in terms of bilinear in $Q$ operators \cite{BGM2013},
\begin{gather}
\langle[\delta \rho(E, \bm{r})]^2\rangle 
=  \frac{\nu^2}{32}\re \Bigl [ P_2(i\varepsilon_{n_1},i\varepsilon_{n_3}) - P_2(i\varepsilon_{n_1} ,i\varepsilon_{n_2}) \Bigl ] \Biggl |_{\footnotesize\hspace{-0.1cm}\begin{array}{l}
\\
i\varepsilon_{n_1,n_3}\to E+i0\\
i\varepsilon_{n_2}\to E-i0
\end{array}
}  ,
 \label{eqK2def0}
\end{gather}
where $\delta\rho(E,\bm{r}) = \rho(E, \bm{r})- \langle\rho(E, \bm{r})\rangle$ and 
 \begin{gather}
 P_2(i\varepsilon_{n},i\varepsilon_{m})
 = \langle\langle  \spp Q_{nn}^{\alpha_1\alpha_1}(\bm{r}) \cdot \spp Q_{mm}^{\alpha_2\alpha_2}(\bm{r}) \rangle \rangle 
 - 2 \langle \spp \bigl [Q_{nm}^{\alpha_1\alpha_2}(\bm{r}) Q_{mn}^{\alpha_2\alpha_1}(\bm{r}) \bigr ] \rangle
 \ .
 \label{eqP2corr}
 \end{gather}
Here $\alpha_{1}\neq \alpha_2$ are some fixed replica indices and $\langle\langle A \cdot B \rangle \rangle = \langle A  B \rangle - \langle A \rangle\langle B \rangle$. Using parametrization \eqref{eq:Q-W} for $Q$ in Eq. \eqref{eqP2corr} and expanding  to the second order in $W$, we obtain
\begin{gather}
P_2(i\varepsilon_{n_1},i\varepsilon_{n_3})
= - \frac{256}{g}\int \frac{d^d \bm{q}}{(2\pi)^d} \Biggl \{
\sin^2 \left (\frac{\theta_{\varepsilon_{n_1}}}{2}\right )\cos^2 \left (\frac{\theta_{\varepsilon_{n_3}}}{2}\right )
\mathcal{D}_q(i\varepsilon_{n_3},-i\varepsilon_{n_1}) 
\notag \\
+ \cos^2 \left (\frac{\theta_{\varepsilon_{n_1}}}{2}\right )\sin^2 \left (\frac{\theta_{\varepsilon_{n_3}}}{2}\right )\mathcal{D}_q(i\varepsilon_{n_1},-i\varepsilon_{n_3})
\Biggr \}  
\end{gather} 
and
\begin{gather}
P_2(i\varepsilon_{n_1},i\varepsilon_{n_2})
= - \frac{256}{g}\int \frac{d^d \bm{q}}{(2\pi)^d} \Biggl \{
\cos^2 \left (\frac{\theta_{\varepsilon_{n_1}}}{2}\right )\cos^2 \left (\frac{\theta_{\varepsilon_{n_2}}}{2}\right )
\mathcal{D}_q(i\varepsilon_{n_1},i\varepsilon_{n_2}) 
\notag \\
+ \sin^2 \left (\frac{\theta_{\varepsilon_{n_1}}}{2}\right )\sin^2 \left (\frac{\theta_{\varepsilon_{n_2}}}{2}\right )
\mathcal{D}_q(-i\varepsilon_{n_2},-i\varepsilon_{n_1})
\Biggr \}   \ .
\end{gather} 
By means of Eqs. \eqref{eq:sol:theta} and \eqref{eq:modif:prop:123}, we find (for arbitrary signs of $\varepsilon$ and $\varepsilon^\prime$)
\begin{equation}
P_2(i\varepsilon,i\varepsilon^\prime)= -\frac{128}{g} \int \frac{d^d \bm{q}}{(2\pi)^d}  
\left [1-  \frac{\varepsilon}{\sqrt{\varepsilon^2+\Delta_{\varepsilon}^2}} \frac{\varepsilon^\prime}{\sqrt{\varepsilon^{\prime 2}+\Delta_{\varepsilon^\prime}^2}}\right ]
\frac{D}{Dq^2+\sqrt{\varepsilon^2+\Delta_{\varepsilon}^2}+\sqrt{\varepsilon^{\prime 2}+\Delta_{\varepsilon^\prime}^2}} \ .
\label{eq:P:final}
\end{equation}
Making an analytic continuation to real frequencies in accordance with the prescription in Eq. \eqref{eqK2def0}, we obtain the following result for the dispersion of fluctuations of the local density of states:
\begin{gather}
\langle[\delta \rho(E, \bm{r})]^2\rangle  =
\frac{4\nu^2}{g} \re \int \frac{d^d \bm{q}}{(2\pi)^d}
\Biggl [
\left ( 1+\frac{E^2}{|\Delta^2(E)-E^2|} \right )
\frac{D}{Dq^2+ 2 \re \sqrt{\Delta^2(E)-E^2}}
\notag \\
- \left ( 1+\frac{E^2}{\Delta^2(E)-E^2} \right )
\frac{D}{Dq^2 +{\color{orange}} 2 \color{black} \sqrt{\Delta^2(E)-E^2}}
\Biggr ] \ .
\label{eq:MLDOS:pert}
\end{gather}
At energies below the gap edge, $E\leqslant E_g$, the gap function is real and satisfy $\Delta(E)\geqslant E$. Therefore, the fluctuations of the local density of states are zero
identically,
\begin{equation}
\langle[\delta \rho(E, \bm{r})]^2\rangle  =
0 \ , \qquad E \leqslant E_g \ .
\label{eq:MesoLDOS:00}
\end{equation}

Above the superconducting gap, $E>E_g$ the mesoscopic fluctuations are non-zero.  At energies close to the gap, $0<E-E_g\ll E_g$, we find
\begin{gather}
\langle[\delta \rho(E, \bm{r})]^2\rangle  \simeq
2 t_0 \rho^2(E) \ln \frac{\min\{L_g,L\}}{\ell} , 
\label{eq:MesoLDOS:0}
\end{gather}
where $L$ stands for the system size and the average density of states, $\rho(E)$  is given by Eq. \eqref{eq:DOS:1}. The length $L_g$ is defined as  
\begin{equation}
L_g = \sqrt{D/(\alpha E_g)} \sim (|\gamma_0|/t_0^2)L_{2\Delta_0} .
\label{eq:def:Lg}
\end{equation} 

Finally, at large energies, $E\gg \Delta_0$, the general expression \eqref{eq:MLDOS:pert}
can be simplified as 
\begin{gather}
\langle[\delta \rho(E, \bm{r})]^2\rangle  \simeq
2 t_0 \rho^2(E) 
\left ( \ln \frac{\min\{L_{*},L_\phi,L\}}{\ell}- \frac{(\re\Delta(E))^2}{2E^2}
\ln \frac{\min\{L_*,L_\phi,L\}}{L_{2\sqrt{E^2-(\re\Delta(E))^2}}}\right ) .
\label{eq:MesoLDOS:1}
\end{gather}
We have neglected $\im \Delta(E)$ everywhere except the denominator of the first diffusive propagator on the right-hand side of Eq. \eqref{eq:MLDOS:pert}. Here, the length $L_*$ is defined through 
\begin{equation}
D/L_*^2 = |\im \Delta^2(E)|/\sqrt{E^2-(\re\Delta(E))^2} .
\label{eq:def:L*}
\end{equation} 
Since $\im \Delta(E)$ is small, the length scale $L_*$ is large, $L_* \gg L_{2\Delta_0}$. In particular,  at $E\sim \Delta_0$ one can find the following estimate: $L_* \sim L_{2\Delta_0} |\gamma_0|/t_0^2 \sim L_g$.
 
We note that the diffusive propagators in Eq. \eqref{eq:MLDOS:pert} are affected by dephasing due to electron-electron interactions \cite{BGM2016}. This results in appearance of the dephasing length $L_\phi$ in Eq. \eqref{eq:MesoLDOS:1}. For energies $E\gg \Delta_0$ the dephasing length can be estimated as 
\begin{equation}
D/L_\phi^2 \sim t(L_E){\color{orange}} \gamma^2(L_E) \color{black} E . 
\label{eq:def:Lphi}
\end{equation}
At the spectral edge, $E=E_g$, $L_\phi$ diverges due to restriction of the phase volume for quasiparticles (there are no quasiparticles below $E_g$). Therefore, for energies close to the spectral edge, $E-E_g\ll E_g$,  the dephasing length exceeds the length $L_g$. This is the reason why $L_\phi$ is absent in Eq. \eqref{eq:MesoLDOS:0}. 

The above results \eqref{eq:MesoLDOS:00}--\eqref{eq:MesoLDOS:1} can be summarized as 
\begin{equation}
\langle[\delta \rho(E, \bm{r})]^2\rangle \simeq 2 t_0 \rho^2(E) \ln \frac{\min\{L,L_*,L_g,L_\phi\}}{\ell} \ .
\label{eq:MesoLDOS:final:pert}
\end{equation}
We remind the reader 
that the lengths $L_g$, $L_*$, and $L_\phi$ are defined in Eqs. \eqref{eq:def:Lg}, \eqref{eq:def:L*}, and \eqref{eq:def:Lphi}, respectively. 
We mention that, with a logarithmic accuracy, $t_0 \ln (\min\{L_*,L_g,L_\phi\}/\ell) \sim 1$ for energies of the order of $\Delta_0$. This indicates that the perturbative treatment of mesoscopic fluctuations has to be extended to a more elaborated renormalization group approach.

\subsection{Renormalization group approach\label{sec:MesoLDOSL:RG}}

The perturbative result \eqref{eq:MesoLDOS:final:pert} has exactly the same form as the perturbative correction to the dispersion of the local density of states above the superconducting transition. Essentially, the superconducting state leads to the specific infrared cut off length scale, $\min\{L_*,L_g, L_\phi\}$, only. This suggests that one can use the one-loop renormalization group equation for $m_2\equiv \langle[\rho(E, \bm{r})]^2\rangle/\rho^2(E)$ derived in Ref. \cite{BGM2016},
\begin{equation}
\frac{d\ln m_2}{d y} = 2 t + O(t^2) \ 
\label{eq:m2}
\end{equation}
upto the length scale $\min\{L_*,L_g,L_\phi\}$. 
Solving Eqs. \eqref{eq:m2} and \eqref{eq:RG:proj:1}, we find
\begin{equation}
\frac{\langle[\rho(E, \bm{r})]^2\rangle}{\rho^2(E)}\simeq \left [ \frac{t(\min\{L,L_*,L_g,L_\phi\})}{t_0}\right ]^2 \ .
\label{eq:RG:MLDOS}
\end{equation}

We note that the length scale $L_*$ starts from the value of the order of $L_g$ and, then, it grows with increasing $E$. The dephasing length $L_\phi$ has the opposite behavior: it decreases with increasing energy. Therefore, there is an energy $E_*$ at which the lengths $L_*$ 
and $L_\phi$ become of the same order. Using 
\begin{equation}
\frac{D}{L_*^2} = \frac{\Delta_0^2}{E} \frac{t_0^2}{|\gamma_0|} u_E^2f(u_E){\color{orange}} |f^\prime(u_E)| \color{black} \ , \qquad 
\frac{D}{L_\phi^2} = \frac{t_0^6}{|\gamma_0|^3} u_E^5 E ,
\label{eq:lengths}
\end{equation}
where $u_E =|\gamma_0|t(L_E)/t_0^2$,
we can find the following estimate, 
\begin{equation}
E_* \sim \Delta_0 {\color{orange}} \left (\frac{|\gamma_0|}{t_0^2}\ln \frac{|\gamma_0|}{t_0^2}\right )^{1/2}\color{black} \ .
\label{eq:E*:def}
\end{equation}
We note that $E_*\gg \Delta_0$. 
Substituting expressions \eqref{eq:lengths} for the length scales $L_*$, and $L_\phi$, into Eq. \eqref{eq:RG:MLDOS}, we obtain the following energy dependence of the  
disorder-averaged second moment of the local density of states (in the case $L=\infty$),
\begin{equation}
\langle[\rho(E, \bm{r})]^2\rangle =
\frac{u_c^2 t_0^2}{\gamma_0^2} \rho^2(E) \begin{cases}
1+ (2 u_c t_0^2/|\gamma_0|)\ln(|\gamma_0|/t_0^2) \ , & \quad E\sim \Delta_0 \ , \\
1- (u_c t_0^2/|\gamma_0|)\ln[\Delta_0 u_E^2 f(u_E) f^\prime(u_E)t_0^2/(E |\gamma_0|)]  \ , & \quad \Delta_0 \ll E\ll E_* \ , \\
\Bigl (1+[u_c t_0^2/(2|\gamma_0|)]\ln [t_0^6 u_E^5 E/(|\gamma_0|^3 \Delta_0)]\Bigr )^{-2}\ , & \quad E_* \ll E\ll 1/\tau \ .
\end{cases}
\label{eq:meso:LDOS:2}
\end{equation}
Interestingly, in accordance with Eq. \eqref{eq:meso:LDOS:2}, the
disorder-averaged normalized second moment of the local density of states has the maximum at $E=E_*$. 
The magnitude of this maximum can be estimated as
\begin{equation}
\frac{\langle[\rho(E_*, \bm{r})]^2\rangle}{\rho^2(E_*)}\frac{\rho^2(E\sim \Delta_0)}{\langle[\rho(E\sim \Delta_0, \bm{r})]^2\rangle} -1
\sim (u_c t_0^2/|\gamma_0|)\ln (|\gamma_0|/t_0^2) \ .
\label{eq:max:est}
\end{equation} 
Since the right hand side of the relation \eqref{eq:max:est} is much smaller than one, we can approximate the disorder-averaged second moment of the local density of states at $E_g<E< E_*$ as 
\begin{equation}
\langle[\rho(E, \bm{r})]^2\rangle = (u_c^2 t_0^2/\gamma_0^2) \rho^2(E) \ .
\end{equation}
This indicates the strong mesoscopic fluctuations of the local density of states at energies $E_g<E< E_*$. 
With further increase of energy, the disorder-averaged second moment of the local density of states decreases. 
Using Eq. \eqref{eq:meso:LDOS:2}, we find  
\begin{equation}
\langle[\rho(E\sim 1/\tau, \bm{r})]^2\rangle \simeq \rho^2(E\sim 1/\tau) \ . 
\end{equation}
Thus, the mesoscopic fluctuations of $\rho(E,\bm{r})$ are suppressed at the ultraviolet energy scale, $E\sim 1/\tau$.

The dependence of $\langle[\rho(E, \bm{r})]^2\rangle$ on the energy is illustrated in Fig. \ref{Fig:DOS}, where the functions $\rho(E)\pm \sqrt{\langle[\rho(E, \bm{r})]^2\rangle}$ are shown by dashed red curves. The fluctuations (not normalized) decreases with increase of the energy. They are strong in the energy interval,  
 $E_g<E<E_*$. At $E_*$ the dependence of $\langle[\rho(E, \bm{r})]^2\rangle$ on the energy is changed, as it was explained above. We note that there is a certain energy $E_{g*}$ which is the solution of the following equation:
$\sqrt{\langle[\rho(E, \bm{r})]^2\rangle}=\rho(E)$. Using Eq. \eqref{eq:meso:LDOS:2}, we find that 
 \begin{equation}
 E_{g*} \sim \left (\frac{\Delta_0}{\tau}\right)^{1/2} \sim \frac{1}{\tau}\exp\left (-\frac{1}{t_0}+\frac{|\gamma_0|}{u_c t_0^2}\right ) \ .
 \label{eq:def:Eg*}
 \end{equation} 
As shown in Fig. \ref{Fig:DOS}, for energies $E<E_{g*}$ the fluctuations of the local density of states are strong. For energies $E>E_{g*}$, the mesoscopic fluctations are not enough to  reduce significantly the local density of states below its average value. 

\subsection{Distribution function for the local density of states \label{sec:MesoLDOSL:typ}}

In a similar way as it was done for the temperatures above $T_c$ \cite{BGM2016}, one can generalize the result \eqref{eq:meso:LDOS:2} to the higher moments of the local density of states, 
\begin{equation}
\langle[\rho(E, \bm{r})]^q\rangle = \rho^q(E) \bigl[\mathcal{A}(E)\bigr ]^{q(q-1)/2} , \qquad \mathcal{A}(E) = \langle[\rho(E, \bm{r})]^2\rangle/\rho^2(E) \ .
\label{eq:LDOS:q}
\end{equation}
We note that according to Eq. \eqref{eq:meso:LDOS:2}, the function $\mathcal{A}(E)>1$ in the energy interval $E_g<E<E_{g*}$, i.e. for region of strong mesoscopic fluctuations. 
The expression \eqref{eq:LDOS:q} implies the following log-normal distribution for the normalized local density of states, $\tilde{\rho}= \rho(E,\bm{r})/\rho(E)$, (see Ref. \cite{Lerner1988} for the case of a normal metal)
\begin{equation}
\mathcal{P}(\tilde\rho) = \frac{\mathcal{A}(E)}{\sqrt{2\pi \ln\mathcal{A}(E)}} \exp\left [ - \frac{1}{2\ln \mathcal{A}(E)} \left (\ln \tilde\rho + \frac{3}{2}\ln \mathcal{A}(E)\right )^2\right ] \ , \qquad E_g<E<E_{g*} \ .
\label{eq:distrib}
\end{equation}
The log-normal distribution \eqref{eq:distrib} predicts that the most probable value for the local density of states is given as
\begin{equation}
\rho_{\rm mode}(E) = \frac{\rho^4(E)}{\langle[\rho(E, \bm{r})]^2\rangle^{3/2}} \ .
\label{eq:rho:mode}
\end{equation}
This result implies that the most probable height of the coherence peak can be estimated as $\rho_{\rm mode}(E\sim \Delta_0)/\nu \sim |\gamma_0|^5/t_0^7$. We note that for $|\gamma_0|\ll t_0\ll |\gamma_0|^{5/7}$ the magnitude of the coherence peak in $\rho_{\rm mode}$ is much larger than $\nu$. In the region $|\gamma_0|^{5/7}\ll t_0\ll |\gamma_0|^{1/2}$ the coherence peak is absent.

We also introduce the typical value of the normalized local density of states, $\tilde{\rho}_{\rm typ}=\exp\langle \ln \tilde\rho\rangle$ where $\langle\dots\rangle$ denotes the average with respect to the distribution \eqref{eq:distrib}. 
Then we obtain,
\begin{equation}
\rho_{\rm typ}(E) = \frac{\rho^2(E)}{\langle[\rho(E, \bm{r})]^2\rangle^{1/2}} \ .
\label{eq:rho:typ}
\end{equation}
This result leads to the following estimate: $\rho_{\rm typ}(E\sim \Delta_0)/\nu \sim |\gamma_0|^3/t_0^5$. 
For $|\gamma_0|\ll t_0\ll |\gamma_0|^{3/5}$, the magnitude of the coherence peak in $\rho_{\rm type}$ is much larger than the bare value of the density of states. For $|\gamma_0|^{3/5}\ll t_0\ll |\gamma_0|^{1/2}$, the coherence peak is absent. We note that, strictly speaking, the typical value of $\tilde{\rho}$ should be determined not from the distribution  \eqref{eq:distrib} but from the modified distribution in which rare events of extremely  small or extremely  large values of $\tilde{\rho}$ are suppressed (see Ref. \cite{Foster2009}). However, as one can check, such a modification does not significantly modify the result \eqref{eq:rho:typ}. 

We mention that the following relation holds: 
\begin{equation}
\rho_{\rm mode}(E)\ll \rho_{\rm typ}(E)\ll \rho(E) \ .
\end{equation}  
This means that for energies $E_g<E<E_{g*}$, an experimentally measured local density of states will be typically  
much smaller than its average value given by Eqs. \eqref{eq:avDOS:final} and \eqref{eq:DOS:1}. 
Nevertheless, for  $|\gamma_0|\ll t_0\ll |\gamma_0|^{3/5}$ a measured local density of states will typically have a high coherence peak and the spectral gap of the order of $E_g$. For $|\gamma_0|^{3/5}\ll t_0 \ll |\gamma_0|^{1/2}$, the coherence peak in a measured local density of states will be completely suppressed and the density of states will have a soft spectral gap of the order of $E_{g*}$. In both ranges of $t_0$, a measured local density of state at some spatial points  will have much higher coherence peak than in the typical spatial regions. For energies $E>E_{g*}$, an experimentally measured local density of states will be close to the disorder-averaged density of states \eqref{eq:avDOS:final}. In this energy interval, $E>E_{g*}$, the mesoscopic fluctuations of the local density of states are small.  This physical picture for the local density of states is illustrated in Fig. \ref{Fig:DOS}.

Following Ref. \cite{BGM2016}, one can also generalize the result \eqref{eq:meso:LDOS:2} to correlation functions of the local density of states at different spatial points and different energies \cite{elsewhere}. Finally, we note that the fact of strong mesoscopic fluctuations of the local density of states is consistent with the strong mesoscopic fluctuations of the superconducting order parameter, see \ref{App:MesoDelta}.

\section{Summary and conclusions\label{s5}}

To summarize, we have developed the theory of the multifractal superconducting state in thin films. Treating the fluctuations around the mean-field spatially homogeneous solution, we derived the modified Usadel equation that incorporates the interplay of disorder and interactions at energy scales larger than the spectral gap. 

\color{black}
Our key findings are as follows:
\begin{itemize}
\item[(i)] The modified Usadel equation, in combination with the self-consistency equation, yields parametrically the same estimate for the multifractally enhanced superconducting transition temperature as the one derived by considering the instability in renormalization group equations in the normal phase. 

\item[(ii)] The mutual effects of disorder and interactions result in strong dependence of the superconducting gap function on energy (see Fig. \ref{Fig:Delta}):  at energies of the order of the spectral gap the gap function $\Delta_{\varepsilon}$ is parametrically enhanced in comparison with its magnitude at ultraviolet energies $\sim 1/\tau$. 

\item[(iii)] The spectral gap at zero temperature is multifractally enhanced in the same way as the superconducting transition temperature. 

\item[(iv)] The energy dependence of the gap function in the Usadel equation results in the profile of the disorder-averaged density of states that, near the spectral gap, resembles the one derived in the model of a spatially random superconducting order parameter (see Fig. \ref{Fig:DOS}). We stress that the interplay of disorder and interactions leads to two opposite effects. On the one hand, it results in the enhancement of the spectral gap, but on the other hand, it induces the effective depairing parameter that 
cuts off
the coherence peaks in the average density of states. The corresponding depairing parameter is estimated as $\sim (t_0^2/\gamma_0)^6 \ll 1$. 

\item[(v)] The mesoscopic fluctuations of the local density of states in the superconducting state are strong at $E<E_{g*}$ (see Eq. \eqref{eq:def:Eg*}). In the energy interval $E_g<E<E_{*}$ (see Fig. \ref{Fig:DOS} and Eq. \eqref{eq:E*:def}) their relative amplitude is of the order of $t_0/|\gamma_0|\gg 1$ that is similar to the estimate in the normal phase at temperatures close to the superconducting transition temperature \cite{BGM2016}. It is worth emphasizing that strong spatial fluctuations of the local density of states (including the local value of the gap and the amplitude of coherence peaks) emerge in our theory despite the fact the model does not involve any macroscopic inhomogeneities. Indeed, our starting point is a model with short-range disorder and with all parameters being spatially uniform. Emergent strong fluctuations are a mesoscopic effect resulting from quantum interference in a disordered system. \end{itemize} 

Based on these findings for the statistics of fluctuations of the local density of states in thin films with multifractally-enhanced superconductivity, we conclude that disorder-induced interference effects dramatically affect spectral properties of these superconducting films, by fully governing the physics in a wide energy interval where 
the spectral gap for single-particle excitations establishes. Specifically, we have identified a parametrically large energy range, $E_g<E<E_{g*}$, where the quasiparticle spectral gap can be zero in some spatial regions and non-zero in the other. The distribution 
function of the local density of states has a log-normal form, such that the typical value of the density of states is lower that the average value. In other words, the system may locally look as superconducting at energies much higher than $E_g$. 

Our results for strong mesoscopic fluctuations of the local density of states are in qualitative agreement with tunneling spectroscopy data on thin superconducting films \cite{Sacepe2008,Sacepe2010,Sacepe2011,Sherman2014,Mondal2011,Noat2013} and with numerical solution of disordered attractive two-dimensional Hubbard model \cite{Fan2020,Stosiek2020}. The strong mesoscopic fluctuations of the local density of states are accompanied by strong mesoscopic fluctuations of the superconducting order parameter.

On the technical side, within our approach, we have integrated over the spatial fluctuations of the superconducting order parameter from the very beginning. Therefore, they are hidden in the term $\hat{S}_{\rm int}^{(c)}$ that describes the interaction in the Cooper channel. The renormalization group analysis of the NLSM in the normal phase suggests that such a procedure is more convenient, since it allows one to take into account the interplay between the spatial fluctuations of the superconducting order parameter (the interaction in the Cooper channel) and charge and spin fluctuations (interaction in the particle-hole channel). This interplay leads to strong renormalization of the NLSM action at length scales of the order of $L_{T_c}$. In our approach, the mesoscopic fluctuations of the local superconducting order parameter manifests themselves in the course of the renormalization group flow as the energy dependence of the gap function.   

Since we considered a macroscopically homogeneous sample, it was natural to address fluctuations around the spatially homogeneous mean-field solution. The emergent fluctuations of the order parameter encoded in the mesoscopic fluctuations of the density of states, as observed in experiments of the ``gap tomography'', are due to mesoscopic quantum interference effects
above the gap-edge energy. At the same time, it is interesting to see whether nonperturbative tails of the density of states inside the gap would appear in spatially homogeneous disordered films studied within the NLSM formalism. For this purpose, one should
consider a possibility of existence of non-trivial saddle-point solutions of the NLSM action. A more straightforward way to obtain such tails would be through the introduction of macrocroscopic inhomogeneities (say, in the local concentration of impurities) that
could induce ``background'' fluctuations of the superconducting gap.

We restricted our consideration to the case of a weak short-ranged interaction
(suppressed by, e.g., high dielectric constant of the substrate), for which a strong effect of the enhancement of superconductivity by multifractality was predicted. 
The approach developed in this paper can be generalized in several directions. Our theory can be extended to include a strong short-ranged interaction, as well as the Coulomb interaction. {\color{orange}} In particular, it would be interesting to study how the multifractality affects the BCS -- BEC crossover \cite{Trivedi2016,Kagan2021}. \color{black} One can also study the multifractal superconducting state that occurs in the system that without superconducting instability is near the interacting metal-insulator transition. Our theory can also be extended to the case of an applied magnetic field that destroys the superconducting state and may produce an intermediate phase with giant magnetoresistance \cite{Trivedi2021} (for studying the giant magnetoresistance near the transition within the NLSM formalism, see Ref. \cite{BGM2015}). Finally, our approach does not take into account phase fluctuations of the order parameter, giving rise to the Berezinskii-Kosterlitz-Thouless phenomena in superconducting films, as well as the existence of the vortices with a normal state in the core. Such fluctuations can be incorporated into our theory in the way similar to the one in Ref. \cite{Konig2015}. We note, however, that for sufficiently high conductance of the normal state,
such effects are expected to influence the results described in this article only slightly.

\section{Acknowledgements}

The authors are grateful to M. Skvortsov and K. Tikhonov for useful discussions. I.S.B. is grateful to M. Stosiek and F. Evers for collaboration on a related project. 
The research was partially supported by the Russian Foundation for Basic Research (grant No. 20-52-12013) -- Deutsche Forschungsgemeinschaft (grant No. EV 30/14-1) cooperation and by the Basic Research Program of HSE.

\appendix

\section{Mesoscopic fluctuations of the superconducting order parameter\label{App:MesoDelta}}

In this Appendix we estimate the mesoscopic fluctuations of the superconducting order parameter. 
While the quantity that is actually measured in experiments on the tomography of the superconductors is the local density of states (Sec. \ref{sec:MesoLDOS}),
the results of the scanning-tunneling-microscopy measurements are frequently translated into the maps of the fluctuating local order parameter by fitting
the density of states with the BCS ``ansatz". Here, instead of translating our results for the mesoscopic fluctuation of the density of states
into the fluctuations of the order parameter, we calculate the latter directly within the NLSM formalism.
The relation \eqref{eq:SCE:full} suggests that the variance of the order parameter can be written as follows:
\begin{equation}
\langle \delta \Delta_r^2(\bm{r}) \rangle = \left (\frac{\pi T\gamma_c}{4}\right )^2
\Biggl [\Bigl \langle\Bigl \langle \Tr \bigl [ t_{r0} L_0^{\alpha_1} Q(\bm{r}) \bigr ]\cdot
\Tr \bigl [t_{r0} L_0^{\alpha_2} Q(\bm{r}) \bigr ]\Bigr \rangle\Bigr \rangle 
- 2 \Bigl \langle \Tr \bigl [ t_{r0} L_0^{\alpha_1} Q(\bm{r}) t_{r0} L_0^{\alpha_2} Q(\bm{r}) \bigr ]\Bigr \rangle \Biggr ]
, 
\label{eq:Var:Delta:def}
\end{equation}
where $\alpha_1\neq\alpha_2$ are some fixed replica indices. We note that the operator in the brackets in Eq. \eqref{eq:Var:Delta:def} is the eigenoperator under renormalization group flow. This can be easily checked with the help of the following identities:
\begin{subequations}
\begin{align}
\langle \Tr A W \Tr B W\rangle & = 2Y \Tr\Bigl [ A B -\Lambda A\Lambda B - A C B^T C+A\Lambda C B^T C \Lambda \Bigr ] , \\
\langle \Tr A W B W\rangle & = 2Y \Bigl [ \Tr A\Tr B - \Tr \Lambda A\Tr \Lambda B +
\Tr A C B^T C - \Tr A\Lambda C B^T C \Lambda\Bigr ], \label{eq:Trans}
\end{align}
\end{subequations}
that follow from  Eq. \eqref{eq:prop:diff}. Here, $Y=(t/2)\ln (L/\ell)$ and we have neglected 
the energy dependence of the diffusive propagators at scales $L\gg L_{T_c}$.

Using parametrization \eqref{eq:Q-W}, expanding the right hand side of Eq. \eqref{eq:Var:Delta:def} to the second order in $W$, and applying Eq. \eqref{eq:prop:diff}, we obtain the following perturbative result:
\begin{gather}
\frac{\langle(\delta \Delta)^2\rangle}{\Delta^2}
= \frac{8}{g} \left ( \sum\limits_{\varepsilon>0} \sin \theta_\varepsilon \right )^{-2}
 \sum\limits_{\varepsilon,\varepsilon^\prime>0}
\sin \theta_\varepsilon \sin \theta_{\varepsilon^\prime} \int \frac{d^d\bm{q}}{(2\pi)^d}
\mathcal{D}_q(i\varepsilon,-i\varepsilon^\prime)  
\label{eq:OP:fluc}
\end{gather}
Here we have used the self-consistency equation \eqref{eq:mod:Usadel:00}. 
We note that the fluctuations $\delta \Delta$ do not lead to a finite single-particle density of states below the gap edge $E_g$ 
but rather correspond to the spatial fluctuations of the superconducting condensate. The hard spectral gap for the quasiparticles results from
an effective averaging over such fluctuations with the self-consistency or renormalization group procedure.\color{black}
With the logarithmic accuracy, we can estimate the right-hand side of Eq. \eqref{eq:OP:fluc} as
\begin{equation}
\frac{\langle(\delta \Delta)^2\rangle}{\Delta^2} \simeq 2 t_0 \ln \frac{\min\{L_T,L_{\Delta_0}\}}{\ell}.
\label{eq:OP:fluc:f}
\end{equation}
Using Eqs. \eqref{eq:Tc:SC} and \eqref{eq:gap:function:final:1}, the
perturbative result \eqref{eq:OP:fluc:f} implies that for temperatures $T\lesssim T_c$ the mesoscopic fluctuations of the superconducting order parameter are large, $\langle(\delta \Delta)^2\rangle \sim \Delta^2$. 
This is consistent with our results from Sec. \ref{subsec:MesoPert}, where the mesoscopic fluctuations of the density of states were calculated, cf. Eq.~(\ref{eq:MesoLDOS:final:pert}). Including renormalization effects as in the calculation of the mesoscopic fluctuations of the density of states in Sec. \ref{sec:MesoLDOSL:RG}, one can generalize Eq.~\eqref{eq:OP:fluc:f} 
in a similar manner, to obtain yet stronger fluctuations of the superconducting order parameter in thin films with multifractally-enhanced superconductivity \cite{elsewhere}.

\section{Kramers--Kronig relations for $\Delta(E)$ \label{App:KK}}

In this appendix, we demonstrate how the Kramers-Kronig relations for $\Delta(E)$,  
\begin{equation}
\im \Delta(E) = - p.v. \int\limits_{-\infty}^\infty \frac{d\omega}{\pi} \ \frac{\re \Delta(\omega)}{\omega-E} \ , \qquad \re \Delta(E) = p.v. \int\limits_{-\infty}^\infty \frac{d\omega}{\pi} \ \frac{\im \Delta(\omega)}{\omega-E} \ .
\end{equation}
are satisfied by the approximate expressions for the real and imaginary parts of $\Delta(E)$.  Let us rewrite  the Kramers-Kronig relation for $\re \Delta(E)$ as
\begin{subequations}
\begin{align}
\re \Delta(E) = p.v. \int\limits_{0}^\infty \frac{d\omega}{\pi \omega} \bigl [ \im \Delta(\omega+E)+\im \Delta(\omega-E) \bigr ]  .
\end{align}
\end{subequations}
Expressing the imaginary part of $\Delta(E)$ in terms of its real part in accordance with Eq. \eqref{eq:Delta:E:Large:Im}, 
\begin{equation}
\im \Delta(\omega) = - (\pi/2) \omega \partial_\omega \re \Delta(\omega) \ ,
\label{eq:an:app}
\end{equation}
we find
\begin{gather}
\re \Delta(E) = - p.v. \int\limits_{0}^\infty \frac{d\omega}{2} \Bigl [ 
\frac{\partial}{\partial \omega} \bigl [\re\Delta(\omega+E)+\re \Delta(\omega-E)\bigr ]
+ \frac{E}{\omega} \frac{\partial}{\partial E} \bigl [ 
\re \Delta(\omega+E)-\re \Delta(\omega-E) \bigr ]\Bigr ] \notag \\
=
\re \Delta(E) + \frac{\pi}{2} E \frac{\partial}{\partial E} \im \Delta(E) .
\end{gather}
We note that here we have neglected the contribution to $\im \Delta(E)$ from energies close to the spectral gap $E_g$. As one can see, when the approximation \eqref{eq:an:app}, which is valid for $E\gg \Delta_0$, is used for all energies,
an additional contribution to the real part of $\Delta(E)$ appears, which is of the order of $(t_0^4/\gamma_0^2) z^2(z^2 f^\prime(z))^\prime$. The latter is of the order of $t_0^4/\gamma_0^2$ at $E\sim \Delta_0$.  This implies that, in order to be able to compute the imaginary part of $\Delta(E)$ from the Kramers-Kronig relation, one needs to derive expression for $\re\Delta(E)$ with the accuracy of the order of $t_0^4/\gamma_0^2$.

\end{document}